\documentclass[12pt]{article}
\usepackage{times}
\usepackage{amsmath}
\usepackage{amsfonts}
\usepackage{graphicx}
\usepackage{ulem}
\usepackage[margin=1in]{geometry}
\usepackage{bm}
\usepackage{hyperref}
\usepackage{cleveref}
\Crefname{figure}{Fig.}{Figs.}
\Crefname{equation}{Eq.}{Eqs.}
\usepackage[font=small,labelfont=bf]{caption}
\usepackage{subcaption}
\usepackage{xcolor}
\Crefformat{equation}{Eq. #2#1#3}

\usepackage{float}
\usepackage[round,numbers,sort&compress]{natbib}
\usepackage{array}
\newcolumntype{L}[1]{>{\raggedright\let\newline\\\arraybackslash\hspace{0pt}}m{#1}}
\newcolumntype{C}[1]{>{\centering\let\newline\\\arraybackslash\hspace{0pt}}m{#1}}
\newcolumntype{R}[1]{>{\raggedleft\let\newline\\\arraybackslash\hspace{0pt}}m{#1}}
\newcommand{\beginsupplement}{%
        \setcounter{section}{0}
        \renewcommand{\thesection}{S\arabic{section}}%
        \setcounter{table}{0}
        \renewcommand{\thetable}{S\arabic{table}}%
        \setcounter{figure}{0}
        \renewcommand{\thefigure}{S\arabic{figure}}%
        \setcounter{equation}{0}
        \renewcommand{\theequation}{S\arabic{equation}}%
        \setcounter{page}{1}
}
\newcommand{\stkout}[1]{\ifmmode\text{\sout{\ensuremath{#1}}}\else\sout{#1}\fi}
\usepackage{titling}

\usepackage[resetlabels]{multibib}
\newcites{Supp}{Supporting References}%

\begin{document}
\title{A versatile framework for simulating the dynamic mechanical structure of cytoskeletal networks}
\author{S. L. Freedman, S. Banerjee, G. M. Hocky, A. R. Dinner}
\date{}
\maketitle
\begin{abstract} 
Computer simulations can aid in understanding how collective materials properties emerge from interactions between simple constituents.
Here, we introduce a coarse-grained model that enables simulation of networks of actin filaments, myosin motors, and crosslinking proteins at biologically relevant time and length scales.
We demonstrate that the model qualitatively and quantitatively captures a suite of trends observed experimentally, including the statistics of filament fluctuations, mechanical responses to shear, motor motilities, and network rearrangements.
We use the simulation to predict the viscoelastic scaling behavior of crosslinked actin networks, characterize the trajectories of actin in a myosin motility assay, and develop order parameters to measure contractility of a simulated actin network. 
The model can thus serve as a platform for interpretation and design of cytoskeletal materials experiments, as well as for further development of simulations incorporating active elements.
\end{abstract}

\pagebreak

\section{Introduction} 

The actin cytoskeleton is a network of proteins that enables cells to control their shapes, exert forces internally and externally, and direct their movements.
Globular actin proteins (G-actin) polymerize into polar filaments (F-actin) that are microns long and nanometers thick.
Many different proteins bind to actin filaments; such proteins often have multiple binding sites that enable them to crosslink actin filaments into networks that can transmit force. 
Myosin proteins are composed of head, neck, and tail domains and aggregate via their tails to form minifilaments that can attach multiple heads to actin filaments \cite{niederman1975}.
Each myosin head can bind to actin and harness the energy from ATP hydrolysis such that a minifilament can walk along an actin filament in a directed fashion---i.e., it is a motor.   
These dynamics have been extensively studied, and it is well understood, for example, how they give rise to muscle contraction.  
In muscle cells, myosin II minifilaments bind to regularly arrayed antiparallel actin filaments and walk toward the barbed ends \cite{huxley1969}. 
In other types of cells lacking this level of network organization, however, the ways in which the elementary molecular dynamics act in concert to give rise to complex cytoskeletal behaviors remain poorly understood.

Addressing this issue requires a combination of experiment, physical theory, and accurate simulation. 
The last of these is our focus here---we present a nonequilibrium molecular dynamics framework that can be used to efficiently explore the structural and dynamical state space of assemblies of semiflexible filaments, molecular motors, and crosslinkers.
By allowing independent manipulation of parameters normally coupled in experiment, this computational model can guide our understanding of the relationship between the microscopic biochemical protein-protein interactions and the macroscopic mechanical functions of assemblies. Additionally, because the model simulates filaments, motors, and crosslinkers explicitly, we can elucidate microscopic mechanisms by studying its stochastic trajectories at levels of detail that are experimentally inaccessible. 
The fact that complex behaviors can emerge from simple interactions also allows simulations to be used to evaluate predictions from theory.

In this work we detail the model and demonstrate that, 
it reproduces an array of known experimental results for actin filaments, assemblies of actin and crosslinkers (passive networks), and assemblies of actin and myosin (active networks). We go further to provide new experimentally testable predictions about these systems.  
For single polymers, we reproduce the spatiotemporal fluctuation statistics of actin filaments.
For passively crosslinked networks, we  reproduce known stress-strain relationships and predict the dependence of the shear modulus on crosslinker stiffness.  
For active networks, we reproduce velocity distributions of actin filaments in myosin motility assays and show how one can tune their dynamical properties by varying experimentally controllable parameters. 
In separate studies, we use the model to clarify microscopic mechanisms of actomyosin contractility and investigate how  assemblies of actin filaments and crosslinkers can be tunably rearranged by myosin motors to form structures with distinct biophysical and mechanical functions \cite{stam2017}.  
The collection of this benchmark suite is itself useful, as prior models \cite{mackintosh1995, head2003, wilhelm2003, kim2009, dasanyake2011, erdmann2012, wang2012, nedelec2007, ennomani2016, gordon2012, kim2014} have focused on specific cytoskeletal features, making the tradeoffs needed to capture the selected behaviors unclear.

Indeed, our model builds on earlier studies, which we briefly review to make clear similarities and differences of the models (see also \cite{popov2016} for a list of cytoskeletal simulations). 
The most finely detailed simulations focus on the motion of a myosin minifilament with respect to a single actin filament.  
Erdmann and Schwarz \cite{erdmann2012} used Monte Carlo simulations to verify a master equation describing the attachment of a minifilament and, in turn, the duty ratio and force velocity curves as functions of the myosin assembly size. 
Stam et al. \cite{stam2015} used simulations to study force buildup on a single filament by a multi-headed motor and found distinct timescale regimes over which different biological motors could exert force and act as crosslinkers. 
These models of actin-myosin interactions are important for understanding the mechanics at the level of a single filament, and their results can be incorporated into larger network simulations. 

A number of publications have dealt with understanding the rheological properties of cross\-linked actin networks \cite{mackintosh1995, head2003, wilhelm2003, kim2009}.  
For example, to study the viscoelasticity of passive networks,  Head et al.\ \cite{head2003} distributed filaments randomly on a two-dimensional (2D) plane, crosslinked filament intersections to form a force-propagating network, sheared this network, and let it relax to an energy minimum.
From this model, they were able to identify three elastic regimes that were characterized by the mean distance between crosslinkers and the temperature. 
Dasanyake and coworkers \cite{dasanyake2011} extended this model to include a term in the potential energy that corresponded to myosin motor activity and observed the emergence of force chains that transmit stress throughout the network. 
These studies address questions about how forces propagate and how crosslinker densities alter mesh stiffness.

Other studies characterized network structure and contractility as functions of model parameters. 
Wang and Wolynes \cite{wang2012} considered a graph of crosslinkers (nodes) and rigid filaments (edges) in which motor activity was simulated via antisymmetric kicks along the filaments. 
They calculated a phase diagram for contractility as a function of crosslinker and motor densities. 
Such a simulation can provide qualitative insights into general principles of filament networks, but the model did not account for filament bending, and structures were sampled via a Monte Carlo scheme that was not calibrated to yield information about dynamics.
Cyron et. al. \cite{cyron2013} used Brownian dynamics simulations to investigate structures that can form via mixtures of semiflexible filaments and crosslinkers and determined a phase diagram and phase transitions \cite{muller2015} between differently bundled actin networks that form as one varies crosslinker density and crosslinker-filament binding angle.
Nedelec and coworkers performed dynamic simulations of assemblies of semiflexible microtubules and kinesin motor proteins, which share features with assemblies of F-actin and myosin; they used their simulation package, CytoSim, to understand aster and network formation in microtubule assays \cite{nedelec2007} and showed recently that the model can be adapted to treat actin networks \cite{ennomani2016}.
Gordon et al. \cite{gordon2012}, Kim \cite{kim2014}, and most recently Popov et al. \cite{popov2016} similarly simulated dynamics of F-actin networks and included semiflexible filaments, motors and crosslinkers. 
By varying motor and crosslinker concentrations, Gordon et al. \cite{gordon2012} and Popov et al. \cite{popov2016} showed various structures that can emerge from assemblies of this type, and Kim \cite{kim2009,kim2014} additionally quantified how these changes could effect force propagation within the network. 

We have strived to include many of the best features of these preceding models in our model.
We use the potential energy of Head et al. \cite{head2003} for filament bending and stretching.
However, in contrast to \cite{head2003, dasanyake2011}, which simply relax the network, we simulate the stochastic dynamics, including thermal fluctuations, crosslinkers and motors binding and unbinding, and the processive activity of myosin.  
The force propagation rules and kinetic equations for binding and unbinding are similar to those of \cite{nedelec2007, gordon2012}, while the length and time scales simulated are on the order of those performed in \cite{kim2014, ennomani2016}.  
We expand on these works by combining and documenting key elements in a single model,  demonstrating that the model can capture experimentally determined trends for cytoskeletal materials quantitatively, and illustrating how the model can be used to study systems of current experimental interest.

\section{Materials and Methods} \label{secn:model}
\begin{figure}[H]  
  \centering
  \includegraphics[width=3.25in]{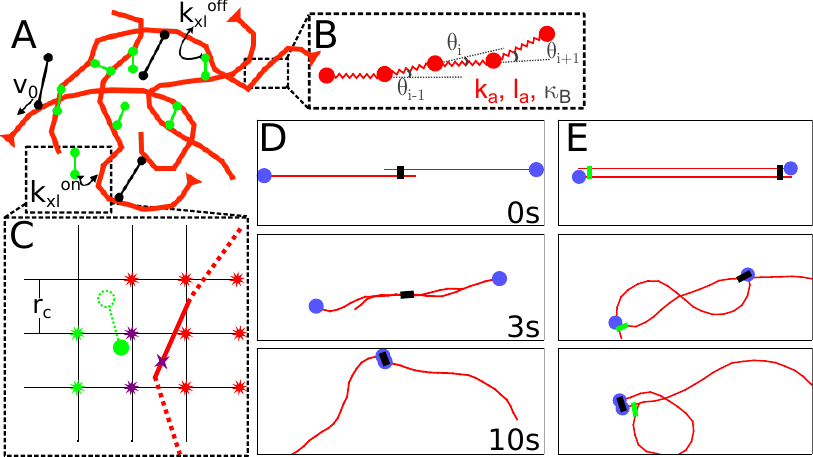}
  \caption{
  \label{fig:toys}%
  Overview of the model.  See \nameref{secn:model} for details. 
  (A) Schematic of a configuration of the model. Filaments are red, crosslinkers are green, and motors are black. 
  (B) Expanded view of the actin filament representation: a chain of beads connected by springs with spring constant $k_a$, rest length $l_a$, and bending modulus $\kappa_B$, as detailed in \nameref{secn:filaments}.
  (C) The process by which a crosslinker finds a filament to bind, as detailed in \nameref{secn:crosslinkers}. 
  The solid red link is indexed to the grid points marked with either red or purple stars, and the solid green motor head searches the grid points marked with either green or purple stars for links to bind. 
  The crosslinker head then stochastically binds to the nearest spot on the filament (see \Cref{app:detbal} and \Cref{fig:detbal}A in the Supporting Materials) here marked with a purple $\times$. 
  (D) Successive images of two antiparallel $10\ \mu$m filaments (barbed end marked in blue) interacting with one motor at the center. The motor binds to both filaments and slides them past each other.
  (E) Similar to (D) but with a crosslinker that pins the top filament's pointed end to the bottom filament's barbed end. The motor, bound to both, walks toward the barbed end of the bottom filament and buckles the top filament.} 
\end{figure}

To access the time and length scales relevant to cytoskeletal network reorganization, we treat actin filaments, myosin minifilaments, and crosslinkers as coarse-grained entities (\Cref{fig:toys}A).  
We model actin filaments as polar worm-like chains (WLC) such that one end of the WLC represents the barbed end of an actin filament and the other represents the pointed end.  
We model crosslinkers as Hookean springs with ends that can bind and unbind from filaments. 
Thus, the connectivity of a network and, in turn, its capacity for force propagation varies during simulations.
We model molecular motors similarly to crosslinkers except that each bound motor head can walk toward the filament barbed end with a load-dependent speed.  
The motors can slide filaments, translocate across filaments, and increase network connectivity.
We simulate the system using Langevin dynamics in 2D because the \textit{in vitro} experiments we wish to interpret are quasi-two-dimensional, and approximating the system as 2D allows us to treat larger systems for longer times. 
To account for the fact that a three-dimensional (3D) system would have greater conformational freedom, we do not include steric interactions for our filaments, motors and crosslinkers. 
This implementation of filaments, motors and crosslinkers, which we detail below, allows for motor-driven filament sliding and filament buckling, as seen in \Cref{fig:toys}D-E.
A complete list of model parameters, their values, and references is provided in \Cref{tab:params}.

\subsection{Filaments}\label{secn:filaments}
The WLC model for actin filaments is implemented as a chain of $N+1$ beads connected by $N$ harmonic springs (links) and $N-1$ angular harmonic springs, as depicted in \Cref{fig:toys}B. 
The $N$ linear springs penalize stretching and keep the filament's average end-to-end length approximately constant.
The $N-1$ angular springs penalize bending and determine the persistence length for a free filament. 
The filament configurations are governed by the potential energy $U_f$:
\begin{equation}
  \begin{aligned}
  U_f &= U_f^{stretch} + U_f^{bend}\\
  U_f^{stretch}&=\frac{k_a}{2}\sum_{i=1}^{N}{(|\vec{r}_i-\vec{r}_{i-1}| - l_a)^2}\\
  U_f^{bend}&=\frac{\kappa_B}{2l_a}\sum_{i=2}^N{\theta_i^2},
  \end{aligned}
  \label{eqn:Ufil} 
\end{equation}
where $\vec{r}_i$ is the position of the $i^{th}$ bead on a filament, $\theta_i$ is the angle between the $i^{th}$ and $(i-1)^{th}$ links, $k_a$ is the stretching force constant, $\kappa_B$ is the bending modulus, and $l_a$ is the equilibrium length of a link. 
In practice, $U_f$ enters the simulation through its Cartesian spatial derivatives (i.e., the forces in \Cref{eqn:overdamped}).  
In this regard, it is important to note that linearized forms for the bending forces are employed in the literature for filaments whose length is constrained via Lagrange multipliers \cite{nedelec2007}, but we found that it was necessary to use the full nonlinear force to obtain consistent estimates for the persistence length, $L_p$, for bead-spring-chain filaments (see \nameref{secn:pl}, below).
We thus employ the full nonlinear Cartesian forces throughout this work, using the expressions in Appendix C of \cite{allen} following the implementation in the LAMMPS Molecular Dynamics Simulator \cite{plimpton1995}.


The bending force constant is derived from the persistence length $L_p$ such that $\kappa_B = L_p k_B T$, where $k_B$ is Boltzmann's constant, and $T$ is the temperature \cite{rubinstein}.  
Experimentally, $L_p = 17\ \mu$m, so $\kappa_B = 0.068$ pN$\mu$m$^2$ for $T = 300$ K \cite{ott1993}.
 
The elasticity per unit length measured for actin filaments with lengths on the order of a micron is $55\pm15$ pN/nm \cite{kojima1994, higuchi1995}.
This implies that a reasonable value for the segment stretching force constant, $k_a$, would be of this order of magnitude.
However, simulating a network of such stiff filaments is computationally infeasible since the maximum timestep of a simulation is inversely proportional to the largest force constant in the simulation \cite{leimkuhler2015}. 
Therefore, we set $k_a$ to a smaller value than estimated from experiment. We note that prevalent extensile behavior, which occurs when filaments interact with two populations of motors with opposite polarities \cite{vale1992},  would necessitate using a more realistic $k_a$.  
However, because $k_a \gg {\kappa_B/l_a^3}$ still, upon compression, filaments prefer bending to stretching, and, as we show, the ability of our model to capture contractile network properties quantitatively is not compromised. 
Unless otherwise indicated, we use $l_a = 1\ \mu$m, because it is the largest segment length that results in the expected spatial and temporal fluctuations for filaments (see \nameref{secn:pl}, below).
We note that very high motor densities can cause filaments to buckle at length scales of $\sim1\ \mu$m, and in these cases it would be necessary to use a smaller $l_a$ to capture those effects \cite{bourdieu1995}. 
 
\subsection{Crosslinkers}\label{secn:crosslinkers}
There are a variety of different actin binding proteins that serve as crosslinkers in the cell cortex, including filamin, fascin, and $\alpha$-actinin. 
Crosslinkers connect filaments dynamically and propagate force within the network.
Thus, the crosslinkers in our model must be able to attach and detach from filaments with realistic kinetic rules and be compliant when bound. 
To this end, we model them as Hookean springs with stiffness $k_{xl}$ and rest length $l_{xl}$. 
Like actin filaments, the Young's modulus of most crosslinkers is significantly higher than would be reasonable to simulate; therefore, for network simulations without large external forces, we set $k_{xl}=k_a$ so that the bending mode of actin filaments is significantly softer than the stretching mode of crosslinkers. 
The rest length $l_{xl}$ corresponds to the size of the crosslinker and therefore differs based on the particular actin binding protein one wishes to study.

The statistics of the bound (\textit{on}) and unbound (\textit{off}) states of each crosslinker are determined by a potential energy of the form
\begin{equation}
  \begin{aligned}
    U_{xl}&=U_{xl}^{stretch}+U_{xl}^{bind}(I_1+I_2)\\
    U_{xl}^{stretch}&=\frac{1}{2}k_{xl}(|\vec{r}_1-\vec{r}_2|-l_{xl})^2\\
    U_{xl}^{bind}&=-k_BT\ln{\left(k_{xl}^{on}/k_{xl}^{off}\right)}
  \end{aligned}
  \label{eqn:uxl}
\end{equation} 
where $\vec{r}_{1 (2)}$ is the position of head $1 (2)$, $I_{1 (2)}$ is $1$ if head $1 (2)$ is bound and $0$ otherwise, and $k_{xl}^{on}$ ($k_{xl}^{off}$) are the rates of binding (unbinding).  

Owing to the form of \Cref{eqn:uxl} and the Monte Carlo rule for binding (below), it is inefficient for a crosslinker to attempt attachment to every filament link in the simulation box.
Rather, we assign a cutoff distance $r_c=\sqrt{k_BT/k_{xl}}$
such that if the distance between a motor and a filament is greater than $r_c$ the probability of attachment is zero. 
This implementation allows us to use the following neighbor list scheme, illustrated in \Cref{fig:toys}C, to determine crosslinker-filament attachment. 
A grid of lattice size of at least $r_c$ is drawn on the 2D plane of the simulation, and each filament link is indexed to the smallest rectangle of grid points that completely enclose it. 
In practice the lattice size is generally larger than $r_c$ due to memory constraints, and is denoted by the model parameter $g$, the number of grid points per $\mu$m in both the $x$ and $y$ directions. 
Since a crosslinker head cannot bind to a filament link that is farther away than $r_c$, it suffices for a crosslinker head to only attempt attachment to the nearby filament links indexed to its four nearest grid points. 


At each timestep of duration $\Delta t$, we enumerate the accessible filament links available to each unbound head.  
For each link, we determine the nearest point to the head's present position and compute a Metropolis factor for moving to that point:  $P_{xl,i}^{off\rightarrow on} = \min[1,\exp(-\Delta U^{stretch}_{xl,i}/k_BT)]$ \cite{metropolis1953}.  
The head then binds to accessible filament link $i$ with probability $(k_{xl}^{on}\Delta t)P_{xl,i}^{off\rightarrow on}$ and stays unbound with probability $1-\sum_i (k_{xl}^{on}\Delta t) P_{xl,i}^{off\rightarrow on}$ \cite{gillespie1977}.  

\sloppy At each timestep, we attempt to move each bound head to a position $\vec{r}_u$ generated by reversing the displacement made upon binding, rotated to account for filament reorientation in the intervening time.  
This choice of $\vec{r}_u$ allows us to satisfy detailed balance for binding and unbinding by accepting the unbinding transition with probability $(k_{xl}^{off}\Delta t)\min[1,\exp(-\Delta U_{xl}^{stretch}/k_BT)]$, as explained in \Cref{app:detbal} in the Supporting Materials.

When both crosslinker heads are attached to filaments, the crosslinker is generally stretched or compressed. 
We propagate the tensile force stored in the crosslinker onto the filaments via the lever rule described in \cite{nedelec2002, gordon2012}. 
Specifically, if the tensile force of a crosslinker head at position $\vec{r}_{xl}$ between filament beads $i$ and $i+1$ is $\vec{F}_{xl}$, then,
\begin{equation}
\begin{aligned}
\vec{F}_i &= \vec{F}_{xl}\frac{|\vec{r}_{i+1}-\vec{r}_{xl}|}{|\vec{r}_{i+1} - \vec{r}_i|}\\
  \vec{F}_{i+1} &= \vec{F}_{xl}-\vec{F}_i 
\end{aligned}
  \label{eqn:lever}
\end{equation}
are the forces on beads $i$ and $i+1$ respectively due to the crosslinker.

\subsection{Motors}\label{secn:motors}
In the present work, we focus on the motor protein myosin II.  As mentioned above, tens of myosin II proteins aggregate into bipolar assemblies called myosin minifilaments
\cite{stam2015}.  
For both myosin minifilaments, and monomeric myosin, motility assay experiments have shown that, on average, bound myosin heads walk toward the barbed end of actin filaments at speeds in the range $0.2-4\ \mu$m/s \cite{kron1986,umemoto1990,harris1993,finer1994}.  
Since myosin also functions to increase the local elasticity of networks where it is bound, we model a motor similarly to a crosslinker, in that it behaves like a Hookean spring with two heads, a stiffness $k_{m}$, and a rest length $l_m$. 
The two heads of this spring do not correspond directly to individual myosin protein heads; rather each of them represents tens of myosin molecules. 
Experimentally minifilaments have a very high Young's modulus, and it is unlikely that their lengths change noticeably in cytoskeletal networks.
As with the passive crosslinkers, we set $k_m=k_a$ so that filament bending is still the softest mode. 
The rest length was set to the average length of minifilaments, $l_m=0.5\ \mu$m \cite{niederman1975}.
Attachment and detachment kinetics, as well as force propagation rules for motors, are the same as for crosslinkers, subscripted with $m$ instead of $xl$ in \Cref{eqn:uxl,eqn:lever}. 

Unlike crosslinkers, motors move towards the barbed end of actin filaments to which they are bound at speeds that decrease with tensile force along the motor.
Myosin motors have been observed to stop walking when the force on them exceeds the stall force, $F_s\approx4$ pN, and most do not step backward \cite{roux2011,riveline1998}. 
We model this behavior by giving each motor head a positive velocity in the direction of the barbed end of the filament to which it is attached; this velocity linearly decreases with the motor's tension projected on the filament, i.e.,
\begin{equation} 
  v(\vec{F}_{m}) = v_0\max\left\{1+\frac{\vec{F}_{m}\cdot{\hat{r}}}{F_s}\ , \ 0\right\},
    \label{eqn:myo_vel}
\end{equation}  
where $v_0$ is the unloaded motor speed, $F_m=-k(|\vec{r}_1-\vec{r}_2|-l_m)$ is the spring force on the motor, and $\hat{r}$ is the tangent to the filament at the point where the motor is bound; $\hat{r}$ points toward the pointed end of the filament.
In the simulations below, we use a value of $v_0=1\ \mu$m/s, which is within range of experimental measurements, but we use a lower value of $F_s=0.5$ pN, so that motors are not stretched to unphysical lengths as they walk.

If the length of a motor's step is larger than the remaining length of filament, then the myosin moves to the barbed end of the filament.
At the barbed end, it has speed $v_0=0$, and detachment rate $k_m^{end}$.
We found that $k_m^{end}=10k_m^{off}$ yielded reasonable results for motility assay and contractile network simulations. 
In experiments, where each myosin minifilament contains many myosins, a lower barbed end affinity may arise from fewer of the minifilament's myosins remaining attached to the actin filament.
In the program, we treat crosslinkers and motors with equivalent objects, but set $v_0 = 0$, and $k_{xl}^{end}=k_{xl}^{off}$ for the crosslinkers.

\subsection{Dynamics}
We use overdamped Langevin dynamics to solve for the motion of filament beads, motors, and crosslinkers.
The Langevin equation of motion for a spherical bead of mass $m$, radius $R$ at position $\vec{r}(t)$ at time $t$, forced by $\vec{F}(\vec{r}(t))$ in a medium with dynamic viscosity $\nu$  is
\begin{equation}
  m\ddot{\vec{r}}(t) = \vec{F}(\vec{r}(t)) + \vec{B}(t) -\dot{\vec{r}}(t)/\mu,
  \label{eqn:lang}
\end{equation}  
where $\vec{B}(t)$ is a Brownian forcing term that introduces thermal energy, and we use the Stokes relation $\mu = (6\pi R\nu)^{-1}$ in the damping term. 
The fastest motions in this simulation are the filament bead fluctuations.  Taking the bead radius to be $0.5\ \mu$m, the maximum speed to be $(2k_BT\mu/\Delta t)^{1/2}=200\ \mu$m/s and the dynamic viscosity to be $\nu = 0.001$ Pa$\cdot$s (corresponding to water), the Reynolds number is very low:  $Re \approx 10^{-4}$. Hence, we treat the dynamics as overdamped and set $m=0$ in \Cref{eqn:lang}.
Furthermore, in the limit of small $\Delta t$, we may write $\dot{\vec{r}}(t) \approx {(\vec{r}(t+\Delta t)-\vec{r}(t))/ \Delta t}$.
These two approximations allow us to rewrite \Cref{eqn:lang} as 
\begin{equation}   
  \vec{r}(t+\Delta t) = \vec{r}(t) + \vec{F}(\vec{r}(t))\mu \Delta t + \vec{B}(t) \mu \Delta t.
  \label{eqn:overdamped}
\end{equation}
For the Brownian term, we use the form of Leimkuhler and Matthews \cite{leimkuhler2013}:
\begin{equation}
  \vec{B}(t)=\sqrt{\frac{2k_BT}{\mu \Delta t}}\left(\frac{\vec{W}(t)+\vec{W}(t-\Delta t)}{2}\right),
  \label{eqn:baoab_brownian}
\end{equation} 
where $\vec{W}(t)$ is a vector of IID random numbers drawn from the standard normal distribution.
This numerical integrator minimizes deviations from canonical averages in harmonic systems; given that all the mechanical forces in our model are harmonic, we expect this choice to yield accurate statistics in the present context as well.
The value for $\Delta t$ in \Cref{eqn:overdamped} is most strongly dependent on the largest force constant in the simulation, $k_a$, but also depends on other simulation parameters  for both motors and crosslinkers, such as $v_0$, $k^{on}$, and $k^{off}$.
\Cref{tab:params} can be used as a rough guide for how high one can set the value of $\Delta t$ for a given set of input parameters; e.g., for a contracting network with $k_a = 1$ pN/$\mu$m, $v_0=1\ \mu$m/s, $k_{xl}^{on}=k_{m}^{on}=k_m^{end}=1$ s$^{-1}$, $k_{xl}^{off}=k_m^{off}=0.1$ s$^{-1}$, and $k_m^{end}=10$ s$^{-1}$, a value of $\Delta t = 2\times10^{-5}$ s is just low enough to iteratively solve \Cref{eqn:overdamped} without accumulating large errors. 

 \subsection{Environment}

In general we use periodic boundary conditions so as to limit finite-size effects. 
We implemented square boundaries to model closed systems, as well as Lees-Edwards boundaries for shearing simulations \cite{allen}. 
The dimensions of the simulation box (\Cref{tab:params}) were chosen to be five times the contour length of filaments so as to be large enough to avoid artifacts due to the self-interaction of constituent components. 

To ignore steric interactions, the fraction $\phi=N_f\pi (D/2)^2 L/V$ of $N_f$ actin filaments (length $L$ and diameter $D$) in a volume $V$ must be lower than the critical volume fraction at which steric interactions yield an isotropic to nematic transition, which for long worm-like chains ($D\ll L_p$ and $D\ll L$) is $\phi_c=5.4D/L$ \cite{prost1995, odijk1986}.
For a network of $500$ filaments of length $L=10\ \mu$m and diameter $D=0.01\ \mu$m, in a $50\times50\times0.1\ \mu$m$^3$ plate, this condition is fulfilled, since $\phi=0.0015<\phi_c=0.0054$.
While it is difficult to estimate the exact thickness of {\it in vitro} experimental actomyosin assays due to the complexity of their preparation, we estimate that they are not thinner than $0.1\ \mu$m \cite{murrell2014methods}. 
We have also ignored hydrodynamic interactions between filament beads; 
the restriction to low packing fraction obviates the need to incorporate anisotropic drag, so we take $\mu$ to be equivalent for both transverse and longitudinal motion \cite{bird1987}.
   
\section{Implementation}
\begin{figure}[H]  
  \centering
  \includegraphics[width=3.25in]{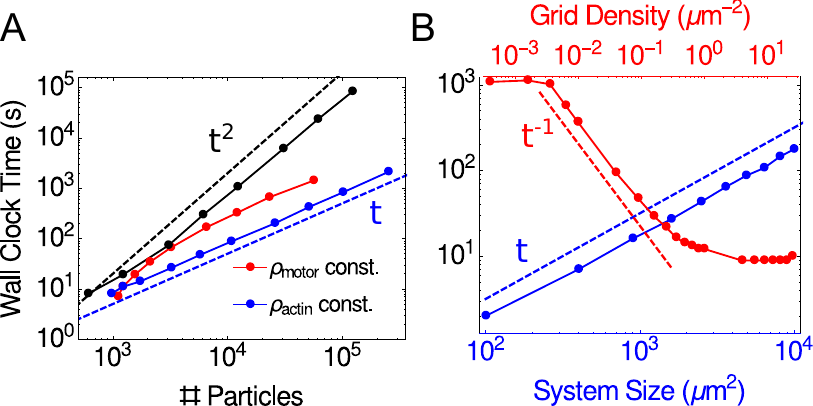}
  \caption{\label{fig:profile}
 Wall clock time for a 10000-step simulation with step size $\Delta t = 0.0001$ s. 
  (A) For a constant system size, run time scales linearly or sublinearly as both filament density (red dots) and motor density (blue dots) are increased independently. 
  If both are increased together (black dots), a quadratic  scaling is approached for large numbers of particles.
  (B) Blue: At constant motor, filament, and grid densities, run time scales linearly with system size (i.e., the area of the simulation box, $XY$). 
  Red: At constant system size, run time decreases with increasing grid density, $g^2$, and thereby the number of neighbor-list grid elements, $g^2XY$, used to calculate motor-filament interactions.
  All benchmarks are for an Intel E5-2680 node with 2 Gb of memory and a 2.70 GHz processor. 
}
\end{figure}

The model is implemented as an open source C++ package called Active Filament Network Simulation (AFiNeS) that is available for download at http://dinner-group.uchicago.edu/downloads.html. 
Installation instructions are available in the README file in the top directory of the AFiNeS package, and all information needed to reproduce the materials in this paper are available in the subfolder ``versatile\_framework\_paper''. 
To run a simulation, a user must compile the code into an executable (e.g., with the provided Makefile) and create an output directory. 
A user can set parameters using command line arguments or a file. 
For example, if the user has compiled the code into the executable ``afines'', created the output directory ``test'', and wants to run a simulation of $500$ $10\ \mu$m actin filaments (with $l_a=1\ \mu$m), interacting with $0.2\ \textrm{motors}/\mu\textrm{m}^2$, and $1\ \textrm{crosslinker}/\mu\textrm{m}^2$ (passive motors), in a cell that is $50\ \mu\textrm{m} \times 50\ \mu\textrm{m}$, for 100 s, he or she could write the following to the file my\_config.cfg
\begin{verbatim} 
xrange=50 # system size in X
yrange=50 # system size in Y

npolymer=500 # number of actin filaments
nmonomer=11 # number of actin beads per filament

a_motor_density=1 # motor density 
p_motor_density=1 # crosslinker density

tf=100 # duration of simulation
dir=``test'' # output directory
 \end{verbatim}
and then run the code using the command 
\begin{verbatim}
afines -c my_config.cfg
\end{verbatim} 
Alternatively, the user could bypass the configuration file and issue the following command:
\begin{verbatim}
afines --xrange 50 --yrange 50 --npolymer 500 --nmonomer 11 \
--a_motor_density 1 --p_motor_density 1 --tf 100 --dir test
\end{verbatim}
In this example, all other parameters were set to their default values (see README file for full list of program parameters).  
With an executable compiled using g++ with the -O3 optimization flag and run on an Intel E5-2680 node with 2 Gb of memory and a 2.7 GHz processor, this example required less than 1.5 days of wall-clock time. 
In general, the wall-clock time of the simulation scales linearly with system size (\Cref{fig:profile}).
\begin{table}[H]
  \caption{Parameter Values}
  \centering
  \begin{tabular}{|C{1.5cm}|L{6.2cm}|C{1.9cm}|C{1.6cm}|C{1.6cm}|C{1.6cm}|}
    \hline\hline
    Symbol & Description (units) (references)
    & \multicolumn{4}{c|}{Simulation}\\ \cline{3-6}
    & & $L_p$ & Shear & Motility Assay & Network \\
    \hline
    &\bf{Actin Filaments}& & & &\\
    \hline
    $N_f$ & Number of filaments & $20$ & $500$ & $1$ & $500$\\
    $N_B$ & Number of beads per filament & $[21,201]$ & $11$ & $[2,26]$ & $11$\\
    $l_a$ & Link rest length ($\mu$m) & $[0.1,1]$ & $1$ & $1$ & $1$\\
    $k_a$ & Stretching force constant (pN$/\mu$m) & $[0.01,1000]$ & $1000$ & $1$ &$1$\\
    $\kappa_B$ & Bending modulus (pN$\mu$m$^2$) \cite{ott1993} & $[0.005,800] $ & $0.068$ & $0.068$ &$0.068$\\
    \hline
    &\bf{Myosin Motors}& & & &\\
    \hline
    $\rho_m$ & Motor density ($\mu$m$^{-2}$) & n/a & n/a & $[0,9]$ & $0.2$ \\
    $l_m$ & Rest length ($\mu$m) \cite{niederman1975} & n/a & n/a & $0.5$ & $0.5$ \\
    $k_m$ & Stiffness (pN$/\mu$m)& n/a & n/a & $1$ & $1$\\
    $k^{on}_m$ & Max attachment rate (s$^{-1}$)& n/a & n/a &$[0.001, 2]$ & $1$\\
    $k^{off}_m$ & Max detachment rate (s$^{-1}$)& n/a & n/a & $1$ & $0.1$\\ 
    $k^{end}_m$ & Max detachment rate at barbed end (s$^{-1}$)& n/a & n/a & $10$ & $1$\\
    $v_0$ & Unloaded speed ($\mu$m/s) \cite{kron1986}&  n/a & n/a & $1$ & $1$\\
    $F_s$ & Stall force of myosin (pN) \cite{veigel2003}& n/a & n/a & $0.5$ & $0.5$\\
    \hline
    &\bf{Crosslinkers} & & & &\\
    \hline
    $\rho_{xl}$ & Crosslink density ($\mu$m$^{-2}$) & n/a & 0.42 & n/a & $1$ \\
    $l_{xl}$& Rest length (Filamin) ($\mu$m) \cite{ferrer2008} & n/a &$0.150$ &n/a & $0.150$\\
    $k_{xl}$ & Stiffness (pN$/\mu$m)& n/a & $[0.1,1000]$ & n/a & $1$ \\
    $k^{on}_{xl}$ & Max attachment rate (s$^{-1}$)& n/a & $1$ & n/a & $1$ \\
    $k^{off}_{xl}$ & Max detachment rate (s$^{-1}$)& n/a & $0.1$ & n/a & $0.1$ \\
    \hline
    &\bf{Environment} & & & &\\
    \hline
    $\Delta t$ & Dynamics timestep (s) &$[10^{-6},10^{-3}]$& $10^{-7}$ & $0.00005$ & $0.00002$\\
    $T_F$& Total simulated time (s) & $2000$ & $0.5$ & $1000$ & $400$\\
    $X$, $Y$ & Length and width of assay ($\mu$m)& n/a & $75$ & $50$ & $50$\\
    $g$ & Grid density ($\mu$m$^{-1}$) & n/a & $2$ & $2$ & $2$\\
    $T$ & Temperature (K)& $300$ & $300$& $300$ & $300$\\
    $\nu$ & Dynamic viscosity (Pa$\cdot$s) & $0.001$& $0.001$& $0.001$ & $0.001$ \\
    $\Delta \gamma$ & Strain (\%) \cite{stricker2010}& n/a& $0.001$&n/a &n/a\\
    $t_{relax}$ & Time between sequential strains (s)& n/a& $0.001$ &n/a & n/a \\
    \hline
  \end{tabular}
  \label{tab:params}
 \end{table} 

\section{Results and Discussion}

In this section, we numerically integrate the model to obtain stochastic trajectories and compare their statistics to known analytical results for semiflexible polymers and networks, as well as experimental observations.  
We also use the model to investigate these systems, including how the viscoelasticity of semiflexible polymer networks depends on crosslinker stiffness, and how the extent of directed motion in actin motility assays depends on filament and motor characteristics.
Finally, we use the model to show how one can quantify contractility in a simulated actin network.

\subsection{Actin filaments exhibit predicted spatial and temporal fluctuations}\label{secn:pl}

The persistence length of a semiflexible filament with bending modulus $\kappa_B$ is expected to be $L_p=\kappa_B/k_B T$.  
However, when simulating the dynamics, approximations can enter both the evaluation of the forces and the discretized numerical integration of the equations of motion.  
Because the persistence length is a measure of filament bending fluctuations, and not an input to the simulation, its dependence on simulation parameters must be determined numerically. 
As discussed in \nameref{secn:filaments} and further below, some care is required to obtain reliable estimates of $L_p$.

\begin{figure}[H] 
  \centering
  \includegraphics[width=3.25in]{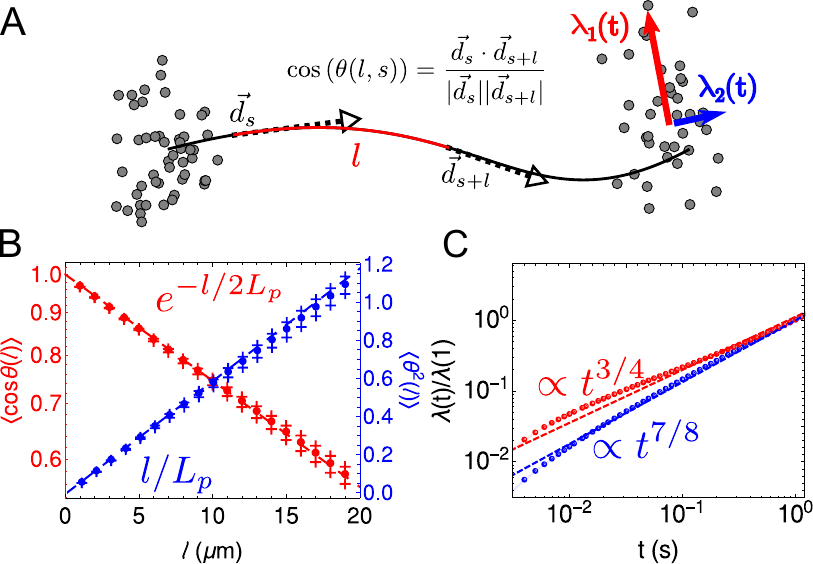}
  \caption{\label{fig:filament}
  Spatial and temporal fluctuations of the bead-spring WLC.
  (A) Schematic of a filament and the order parameters that characterize its fluctuations. 
  Spatial fluctuations are characterized by the angle between two tangent vectors $\vec{d}_s$ and $\vec{d}_{s+l}$ along the filament as a function of the contour length between them, $l$. 
  Temporal fluctuations are characterized by the eigenvalues $\lambda_{1,2}(t)$ of the covariance matrix of filament endpoint positions as a function of time. 
  The red arrow indicates the larger moment ($\lambda_1$, measuring transverse fluctuations) while the blue arrow indicates the smaller moment ($\lambda_2$, measuring longitudinal fluctuations).
  (B) Decorrelation of tangent vectors (red circles) and fluctuations in angles between links (blue circles) as a function of the arc length between them. 
  For the $N=20$ filaments analyzed, the blue (red) dots show the mean of $\langle\theta(l)^2\rangle$ $\left(\langle\cos{(\theta(l))}\rangle\right)$ and the error bars show their standard errors, $\sigma/\sqrt{N}$, where $\sigma$ is their standard deviation.    
  Dashed lines show expected behavior for $\kappa_B=0.068$ pN$\mu$m$^2$.
  (C) Eigenvalues of covariance matrices for the positions of endpoints of filaments as a function of time.
  Red dots show $\lambda_1(t)$, which is expected to be proportional to $t^{3/4}$ (red line) while blue dots show $\lambda_2(t)$, which is expected to be proportional to $t^{7/8}$ (blue line). 
  Standard error is smaller than the size of the data points.
}
\end{figure}

For a two dimensional filament it is possible to show analytically that if a small bend between links $i$ and $i-1$ of an $N$ link chain results in a local change in free energy of ${(\kappa_B/2l_a)\theta_i^2}$, then 
\begin{equation}
  \langle\theta^2(l)\rangle = {l/ L_p}
  \label{eqn:thsq}
\end{equation}
\begin{equation} 
  \langle\cos(\theta(l))\rangle = \exp{(-l/2L_p)},
  \label{eqn:costh}
\end{equation} 
where $\theta(l) = \theta_j - \theta_i$, $l = l_a(j-i)$ $(2\le i<j\le N)$ \cite{frontali1979}.  
To test our WLC model against these predictions, we let $20$ filaments of $L=20\ \mu$m and $\kappa_B=0.068$ pN$\mu$m$^2$ fluctuate at $T=300$ K for $T_f = 2000$ s and measured the resulting filament configurations.
The configurations saved were chosen to be 2 s apart, since the decorrelation time for $\theta(l)$ was at most 1.1 s (see \Cref{app:fil} and \Cref{fig:pl_supp_decor} for details). 
The first $100$ s of each simulation was disregarded as filaments had not yet equilibrated.
For each of the $20$ filaments, we evaluated $\langle\theta^2(l)\rangle$ and $\langle\cos(\theta(l))\rangle$ for each $l\in{1,2,\dots,19}\ \mu$m from its $1900$ saved configurations. 
We show the average for each of these values over all filaments in \Cref{fig:filament}B, along with the expected behavior, given the input $\kappa_B$. 

\begin{figure}[H]
  \centering
  \includegraphics[width=3.25in]{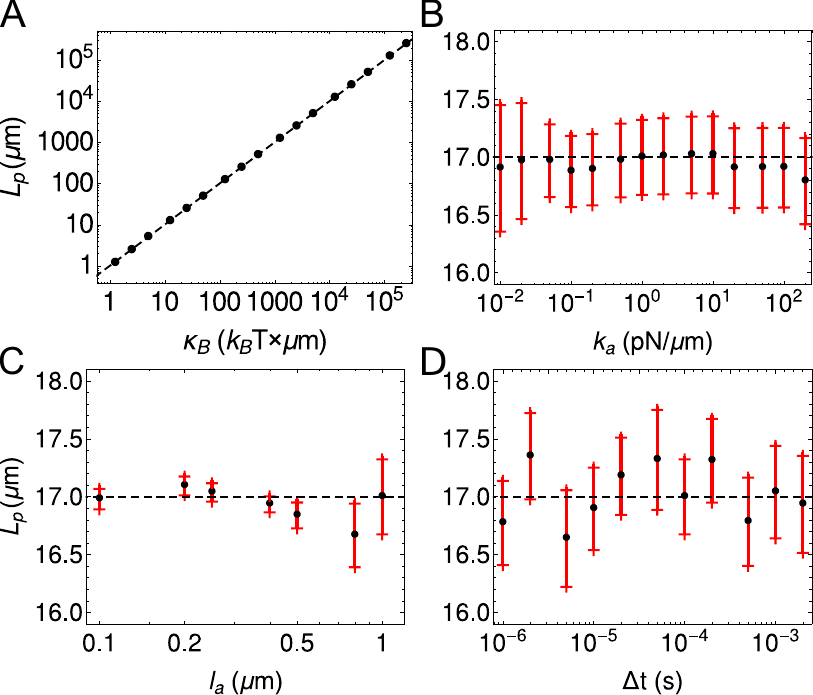}
  \caption{
   \label{fig:pl_supp}
   Dependence of the persistence length on the parameters for numerically integrated semiflexible filaments. 
   Error bars are $\sigma/\sqrt{N}$, where $\sigma$ is the standard deviation of the values of $L_p$ obtained from fitting a line to the first $5$ data points of $\langle\theta^2(l)\rangle$ for each of the $N=20$ filaments simulated.  
   The dashed lines show the predicted persistence length, based on the input bending modulus $\kappa_B$. 
   The default parameters are $\kappa_B = 17\mu$m$\times k_B T$, $k_a = 1$ pN$/\mu$m, and $l_a=1\mu$m. 
   In (A)-(C), $\Delta t\ge10^{-6}$ s, and the largest $\Delta t$ that yielded stable integration was used.   
   In (C), $k_a = 1$pN$/l_a$.
 }
 
 \end{figure}

As alluded to above, the numerical integration can make the persistence length depend on simulation parameters in nonobvious ways.  Consequently, we measured the sensitivity of $L_p$ to independent variations of $\kappa_B$, $l_a$, and $k_a$.  
The results shown in \Cref{fig:pl_supp} are obtained from using the definition $L_p = 1/({d\langle\theta^2(l)\rangle/dl})$ (i.e., the inverse of the slope of the ``blue'' line in \ref{fig:filament}B).
\Cref{fig:pl_supp}A shows that in the range of  $\kappa_B \in [1,10^5]\ \mu$m$\times k_B T$, $L_p$ determined from the simulation agrees well with the input bending modulus, and can be easily tuned to simulate filaments of varying rigidity. 
\Cref{fig:pl_supp}B shows that for a wide range of link stiffnesses, $L_p$ is  independent of $k_a$. 
We also tested the dependence of $L_p$ on the link rest length, $l_a$. 
In thermal equilibrium, the variance of the link lengths is $\langle\Delta l_a^2\rangle=k_BT/k_a$. 
Thus, to keep the fluctuations in the filament's contour length $L$ constant, one should set $k_a\propto l_a^{-2}$. 
In practice, this scaling is computationally difficult to achieve  when $l_a<0.3\ \mu\mathrm{m}$ because  high $k_a$ requires a very small $\Delta t$ in \Cref{eqn:overdamped}.
We therefore used a less steep variation, $k_a=1$ pN$/l_a$, and show in \Cref{fig:pl_supp}C that consistent values of $L_p$ are obtained when $l_a\in[0.1,1]\ \mu$m.  
We thus see that, there is a range in which $L_p$ is independent of the filament link parameters, $k_a$ and $l_a$, although high stiffness and low link length both require using a small timestep, and therefore limit the duration of the simulation. 
In \Cref{app:cytosim} (\Cref{fig:cs_lp}) we measure the persistence length of fibers simulated using Cytosim and obtain similar results. 

The statistics of temporal fluctuations are also known for semiflexible filaments.
Fluctuations transverse to the filament orientation increase as $\langle dr_{\perp}^2\rangle\propto t^{3/4}$, while longitudinal fluctuations increase as $\langle dr_{||}^2\rangle\propto t^{7/8}$ \cite{everaers1999}. 
To determine if our simulations agreed with these theoretical scaling relations, we followed the procedure outlined in \cite{everaers1999} and generated $N = 100$ initial filament configurations of a $20\ \mu$m filament. 
This length was chosen because it satisfied the constraint provided in \cite{everaers1999} for the fluctuations of the two ends of the filament to be uncorrelated at long times (here $t=1$ s); i.e., $20\ \mu$m $ > \left({tk_BT/\nu}\right)^{1/8} \left({\kappa_B/k_B T}\right)^{5/8}=7\ \mu\textrm{m}$.  
For each configuration we ran $M = 1000$ simulations of the filament diffusing freely for $1$ s.  
We denote each of the $M$ positions for each endpoint at each time by $\vec{r}_e(t)$. 
For each of the clouds of points shown in \Cref{fig:filament}A, we calculated the moments, as the eigenvalues of the covariance matrix  with elements $\langle (\vec{r}_e(t)\cdot \hat{i}-\langle \vec{r}_e(t)\cdot \hat{i}\rangle)(\vec{r}_e(t)\cdot \hat{j}-\langle \vec{r}_e(t)\cdot \hat{j}\rangle)\rangle$ for $i,j\in\{x,y\}$.
The larger eigenvalue $\lambda_1(t)$ corresponds to the transverse fluctuations (i.e., $\lambda_1(t)\propto t^{3/4}$) while the smaller eigenvalue corresponds to the longitudinal fluctuations ($\lambda_2(t)\propto t^{7/8}$). 
We show these results in \Cref{fig:filament}C. 
Each data point is the average over the $2NM$ eigenvalues for $\lambda_1(t)$ and $\lambda_2(t)$.
As evident, the computed scaling relations are in good agreement with theoretically predicted behaviors. 
  
\subsection{Tunable elastic behavior of crosslinked filament networks}\label{secn:shear}
\par
The mechanical properties of crosslinked F-actin have important ramifications for force generation and propagation within a cell. They are generally inferred from rheological measurements of \textit{in vitro} networks \cite{gardel2004,koenderink2006,kasza2009,lin2010}.
In a typical experiment, actin and crosslinker proteins are mixed to form a crosslinked mesh and then sheared in a rheometer by a prestress, $\sigma_0$. 
The prestressed network then undergoes a sinusoidal differential stress of magnitude $d\sigma\ll\sigma_0$. 
By measuring the resulting strain, one can calculate the differential elastic modulus $G(\sigma_0) = {d\sigma/ d\gamma}$.  
Results from such experiments indicate that, in contrast to a purely viscous fluid, crosslinked F-actin networks resist shear, and $G$ increases nonlinearly with stress indicative of shear stiffening.

In experiments using a stiff crosslinker, such as scruin, the dependence of the differential modulus on high prestress is $G\propto\sigma_0^{3/2}$ \cite{gardel2004,lin2010}. 
Force-extension experiments with semiflexible filaments, in which one directly measures the force $F$ required to extend a filament by a distance $l$, yield a remarkably similar relationship, $dF/dl\propto F^{3/2}$ \cite{bustamante1994,marko1995}.
As remarked in \cite{gardel2004}, this suggests that the shear stiffening is a direct result of the nonlinear force-extension relationship of actin. 
Rheology studies using more compliant crosslinkers, such as filamin, have found a softer response, $G\propto\sigma_0$, indicating that a significant amount of stress is mediated through the crosslinkers, and not the filaments \cite{kasza2009}.  
These results suggest that the strain stiffening behavior of a crosslinked network can be tuned by varying the crosslinker stiffness. 

To test this possibility and benchmark our simulations, we subjected passive networks comprised of filaments and crosslinkers to shear.  
We initialized each simulation with $N = 500$ randomly oriented filaments of length $15\ \mu$m in a square box of area $75\ \mu\textrm{m} \times 75\ \mu\textrm{m}$. 
A $0.150\ \mu$m crosslink (corresponding to the length of filamin) was initially placed at each filament intersection.  
To inhibit network restructuring, the detachment rate of the crosslinkers was set to zero. 
We performed 24 such simulations, each with a different crosslinker stiffness in the range $0.1-1000$ pN/$\mu$m.  

Simulating shear rheology experiments requires modifying the equations of motion and the boundary conditions to achieve a planar Couette flow.
In general, planar Couette flow can be simulated via molecular dynamics using 
Eq. (4.1) in \cite{evans1984}:
\begin{equation}
 \begin{aligned}
  m\ddot{x}&=F_{int,x}+\dot{\gamma}y\\
  m\ddot{y}&=F_{int,y},
  \end{aligned}
  \label{eqn:evans2}
\end{equation}
where $x$ and $y$ are the Cartesian coordinates of a particle being sheared, $F_{int,x}$, $F_{int,y}$ are the internal forces on those particles and $\gamma$ is the strain. 
Simultaneously, the upper and lower boundaries must be sheared by the total strain on the simulation box \cite{allen}.
Comparing \Cref{eqn:evans2} with \Cref{eqn:lang}, we substitute $F_{int,x} = F(x(t)) + B_x(t) - \dot{x}(t)/\mu$. 
In the overdamped limit, $\ddot{x}_i=0$, so implementing \Cref{eqn:evans2} is equivalent to updating filament bead positions via \Cref{eqn:overdamped}, and shifting the horizontal position of a bead ($x_i$) by 
\begin{equation}
  x_i \rightarrow x_i + \Delta\gamma \left( \frac{y_i}{Y} \right),
  \label{eqn:sllod}
\end{equation}
where $\Delta \gamma=\dot{\gamma}\Delta t $ and $Y$ is the simulation cell height. 
The boundary conditions follow the Lees-Edwards convention \cite{allen}. 

Since moving the particles $\Delta\gamma$ is equivalent to the addition of a significant external force on the system, it is necessary to let the network relax for a specified amount of time $t_{relax}$ after each shear event, before measuring the network's internal energy. 
The magnitude of $t_{relax}$ depends on $\Delta \gamma$, which in turn depends on the chosen discretization of the strain and the timestep $\Delta t$. 
As shown in \Cref{app:shear} and \Cref{fig:elas_supp}, we found that $\Delta \gamma=0.001$, $\Delta t=10^{-7}$ s, and $t_{relax} =0.001$ s yielded a stable planar Couette flow, with high enough strains to observe strain stiffening.
This protocol was performed for $T_f=0.5$ s yielding a total strain of $\gamma=\Delta \gamma T_f/t_{relax}=0.5$. 

\begin{figure}[H]
  \centering
  \includegraphics[width=3.25in]{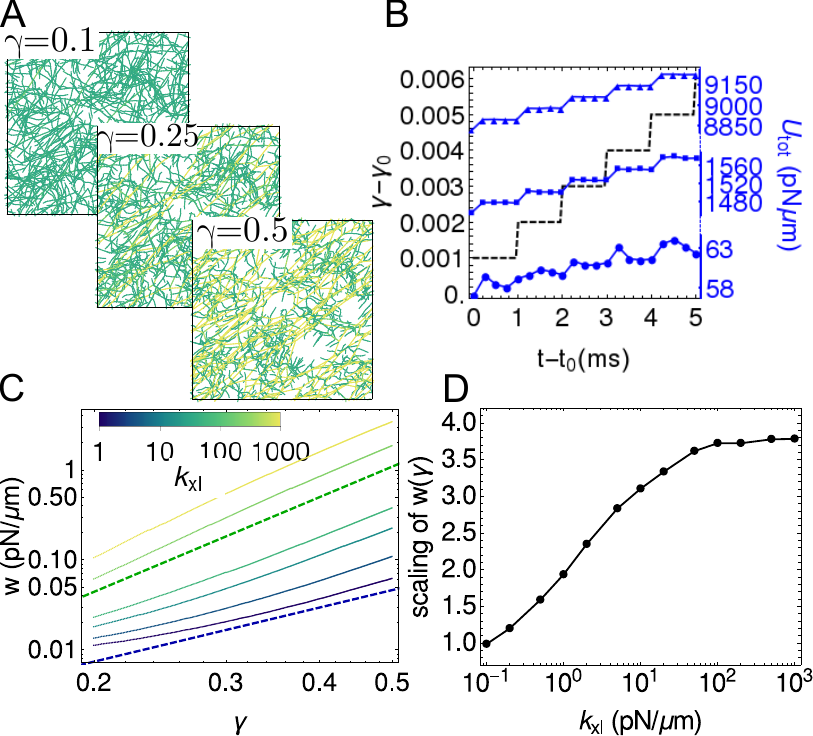}
  \caption{%
    \label{fig:stress}%
    Tunable elasticity of crosslinked networks. 
    (A) Snapshots of a strained network ($k_a=1000$ pN$/\mu$m, $k_{xl}=20$pN$/\mu$m) at $\gamma=0.1$, $\gamma=0.25$, and $\gamma=0.5$. 
    Color indicates stretching energy on each link, with green being the lowest and yellow being the highest. 
    For all snapshots, $t=\gamma \times 1$ s.  
    (B) The potential energy of the network as a function of time shown at different strains $\gamma_0=0.1$ (circles), $\gamma_0=0.25$ (squares), and $\gamma_0=0.4$ (triangles), where $t_0=\gamma_0\times 1$ s. 
    Black dashed line shows the strain protocol.   
    (C) Strain energy density ($w=U/XY$) for various values of crosslinker stiffness $k_{xl}$. 
    Blue dashed line indicates expected behavior for a linearly elastic solid ($w\propto \gamma^2$) and green dashed line indicates strain stiffening behavior of $w\propto \gamma^{3.5}$ as expected for semiflexible polymer networks \cite{gardel2004,lin2010}.
   (D) The power-law exponent of $w(\gamma)$ as a function of crosslinker stiffness, evaluated by least squares fitting $\ln{(w)}$ as a function of $\ln{(\gamma)}$. 
  }
\end{figure}

We measured the elastic behavior of the network for each crosslinker stiffness by calculating $w$, the strain energy density at each timestep:
\begin{equation}
  w(t) = \frac{1}{X Y}\left(\sum_f{ U_f}+\sum_{xl}{U_{xl}}\right),
  \label{eqn:sed}
\end{equation}
where $U_f$ is the mechanical energy of individual filaments (\Cref{eqn:Ufil}) and $U_{xl}$ is the mechanical energy of each crosslink (\Cref{eqn:uxl}). 
By averaging over windows of size $t_{relax}$, we determine $w(\gamma)$. 
\Cref{fig:stress} shows the results of these calculations for various values of $k_{xl}$. 
For extremely low $k_{xl}$, the strain energy scaled linearly with strain, $w\propto \gamma$, indicating that the network showed no resistance to shear: $G={d^2w/d\gamma^2}=0$. 
For high $k_{xl}$, we observe a neo-Hookean strain stiffening behavior, $w\propto\gamma^4$ \cite{shokef2012}. 
Thus, we can tune the material properties of crosslinked semiflexible networks from being liquid-like, with $w\propto\gamma$, through the Hookean elastic regime of $w\propto\gamma^2$ up to the strain stiffening regimes of $w\propto\gamma^3$ and $w\propto\gamma^{3.5}$, as previously reported in experiments \cite{gardel2004, kasza2009}. 
We show in \Cref{app:sheareng} and \Cref{fig:engs} that the monotonic increase in scaling for low $k_{xl}$ corresponds to a regime where the strain energy is mostly stored in the crosslinkers, while the plateau at high $k_{xl}$ corresponds to a regime where the strain energy is mostly stored in filaments.  

\subsection{Ensembles of motors interacting with individual filaments simulate actin motility assays}\label{secn:motility}
While the attachment, detachment and speed of an individual myosin motor is a model input (described in \nameref{secn:crosslinkers} and \nameref{secn:motors}, above), the collective action of many motors on a filament is an output that can be compared with actin motility assays \cite{riveline1998, walcott2012}. 
In the canonical motility assay experiments, a layer of myosin is attached to a glass coverslip, and actin filaments are distributed on top of the layer of myosin motors. 
The fixed motors translocate the actin filaments. 
The speed of an actin filament has been reported to depend nonlinearly on the concentration of myosin and the concentration of ATP in the sample \cite{harris1993, umemoto1990}. 
Thus, by allowing the filaments to interact with more motors, one can monotonically increase the filament speed to a constant value.

\begin{figure}[H] 
    \centering
   \includegraphics[width=6.75in]{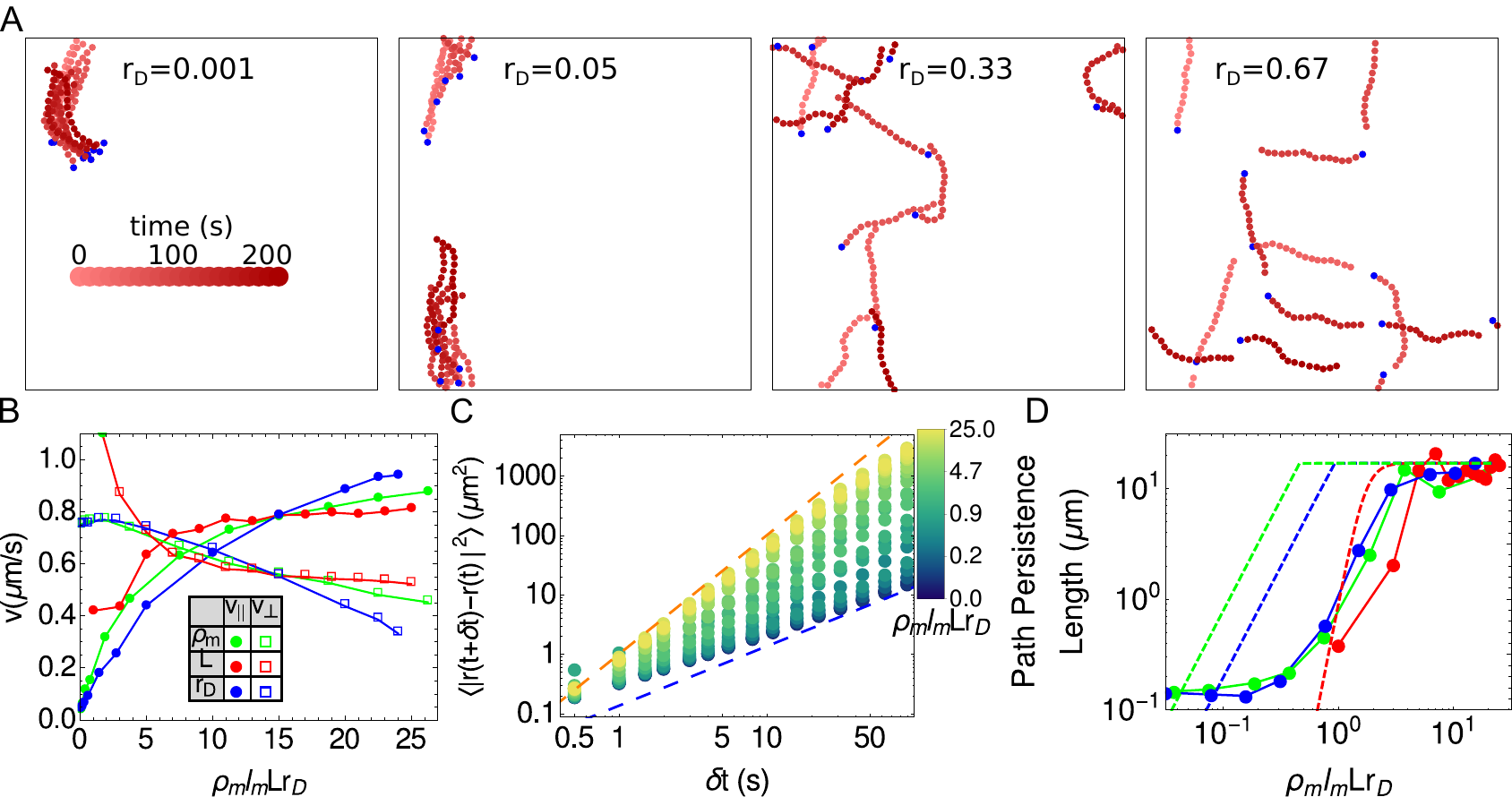}
  \caption{%
  \label{fig:motility}%
  Nonlinear dependence of filament motility on motor-filament interaction probability.
  (A) Trajectory of a filament for $\rho_m = 4\ \mu\textrm{m}^{-2}$ and $L = 15\ \mu$m as a function of time for different values of the duty ratio, $r_D$. 
  Depth of color indicates time of the snapshot, as indicated by the scale. 
  Blue dot marks the barbed end.  
  (B) Filament speed decomposed into longitudinal (filled circles) and transverse (empty squares) components as a function of the dimensionless parameter $\mathcal{M} = \rho_ml_mLr_D$,
  by independently varying $\rho$ (green), $L$ (red), and $r_D$ (blue).  
  The default parameters were $\rho=4\ \mu$m$^{-2}$, $L = 15\ \mu$m, and $r_D=0.5$. 
  (C) Mean squared displacement for various values of $\mathcal{M}$.  
  Blue dashed line shows diffusive behavior and orange dashed line shows ballistic behavior.
  (D) Path persistence length for simulations described in (B), evaluated via \Cref{eqn:costh} over $5$ replicates. 
  Dashed lines are theoretical predictions for these values using equations (1)-(6) in \cite{duke1995}.
 }
\end{figure}

To explore the dynamics of such an assay, we randomly distributed motors on a $50\ \mu$m $\times 50\ \mu$m periodic simulation cell and tethered one head of each motor to its initial position. 
These model motors represent myosin minifilaments with dozens of heads, and therefore have a high default duty ratio ($r_D=0.5$), and rest length $l_m=0.5\ \mu$m \cite{niederman1975,ideses2013}.
Filaments were then introduced in the simulation cell and allowed to interact with the free motor heads. 
The strength of motor-filament interactions was manipulated in three ways: by varying the motor concentration $\rho_m$, the filament contour length $L$, and the duty ratio $r_D =k_m^{on}/(k_m^{on}+k_m^{off})$. 
While $L$ and $r_D$ are difficult to modulate experimentally in a well-controlled fashion, as they require the addition of other actin-binding proteins to the assay,  they are predicted to impact the dynamics of actin by varying the number of myosin heads bound to an actin filament at any one time \cite{harris1993}. 
Since they are both simple functions of the model's parameters, we were able to test this hypothesis directly.
We plot our simulation results in \Cref{fig:motility} as functions of the dimensionless control parameter $\mathcal{M} = \rho_m l_m L r_D $ (where $\rho_ml_m$ is the linear motor density), which represents the average number of bound motor heads per filament.

Our findings are qualitatively similar to the previously reported experimental results and expand on them by collapsing the trends observed while varying $\rho_m$, $L$, and $r_D$ into a single effective parameter. 
At low $\mathcal{M}$, i.e., low motor density, filament length, or duty ratio, \Cref{fig:motility}B shows that transverse motion dominates over longitudinal motion as the filament is not propelled by motors faster than diffusion, and transverse filament fluctuations are larger than longitudinal fluctuations (consistent with \Cref{fig:filament}C). 
However, as $\mathcal{M}$ increases, longitudinal motion dominates. 
Consistent with  experimental results \cite{kron1986,harris1993}, the longitudinal speed of the filament plateaus at
$v_{||}\approx 1\ \mu$m/s, which is the input unloaded speed of a single motor.
In \Cref{fig:motility}C, we plot the mean squared
displacement (MSD) of the filament,  $\langle{\Delta r}^2\rangle=\langle|\vec{r}(t+\delta t)-\vec{r}(t)|^2\rangle$ with angle brackets indicating an average over time $t$.
We show that low $\mathcal{M}$ yields diffusive behavior with $\langle {\Delta r}^2 \rangle \propto \delta t$, and high $\mathcal{M}$ yields ballistic  motion with $\langle {\Delta r}^2 \rangle\propto {\delta t}^2$. 
We obtain similar results for motility assay behavior with a corresponding Cytosim simulation, as shown in \Cref{app:cytosim_mot} (\Cref{fig:cs_motility}).

An interesting outcome of these simulations is how the direction of a filament changes over time for varying $\mathcal{M}$.  Specifically, we calculate the directional autocorrelation of a filament and, in turn,  the persistence length of the path of the filament by applying \Cref{eqn:costh} to the center of mass of the filament at frames separated by $\Delta t=1$ s (\Cref{fig:motility}D).  
Scaling arguments suggest that the path's persistence length depends strongly on motor density, duty ratio, and filament length \cite{duke1995}. 
In the limit of high $\mathcal{M}$, the distance between motors that are bound to a filament is sufficiently short that the filament does not diffuse transversely; however, fluctuations in the filament configuration still allow directional decorrelation, and consequently the path's persistence length is $L_p$. 
At low $\mathcal{M}$, the distance between bound motors is sufficiently large that rotational diffusion causes the filament's path to be completely decorrelated, such that the path's persistence length approaches $0$. 
In \Cref{fig:motility}D, we show that the simulation agrees with theoretically predicted scaling laws at low and high $\mathcal{M}$ \cite{duke1995}.  
Our results delineate the values of $\mathcal{M}$ at which there are crossovers between the predicted limiting regimes.

\subsection{Molecular motors cause flexible, crosslinked networks to contract}\label{secn:contract}
\begin{figure}[H] 
    \centering
    \includegraphics[width=6.75in]{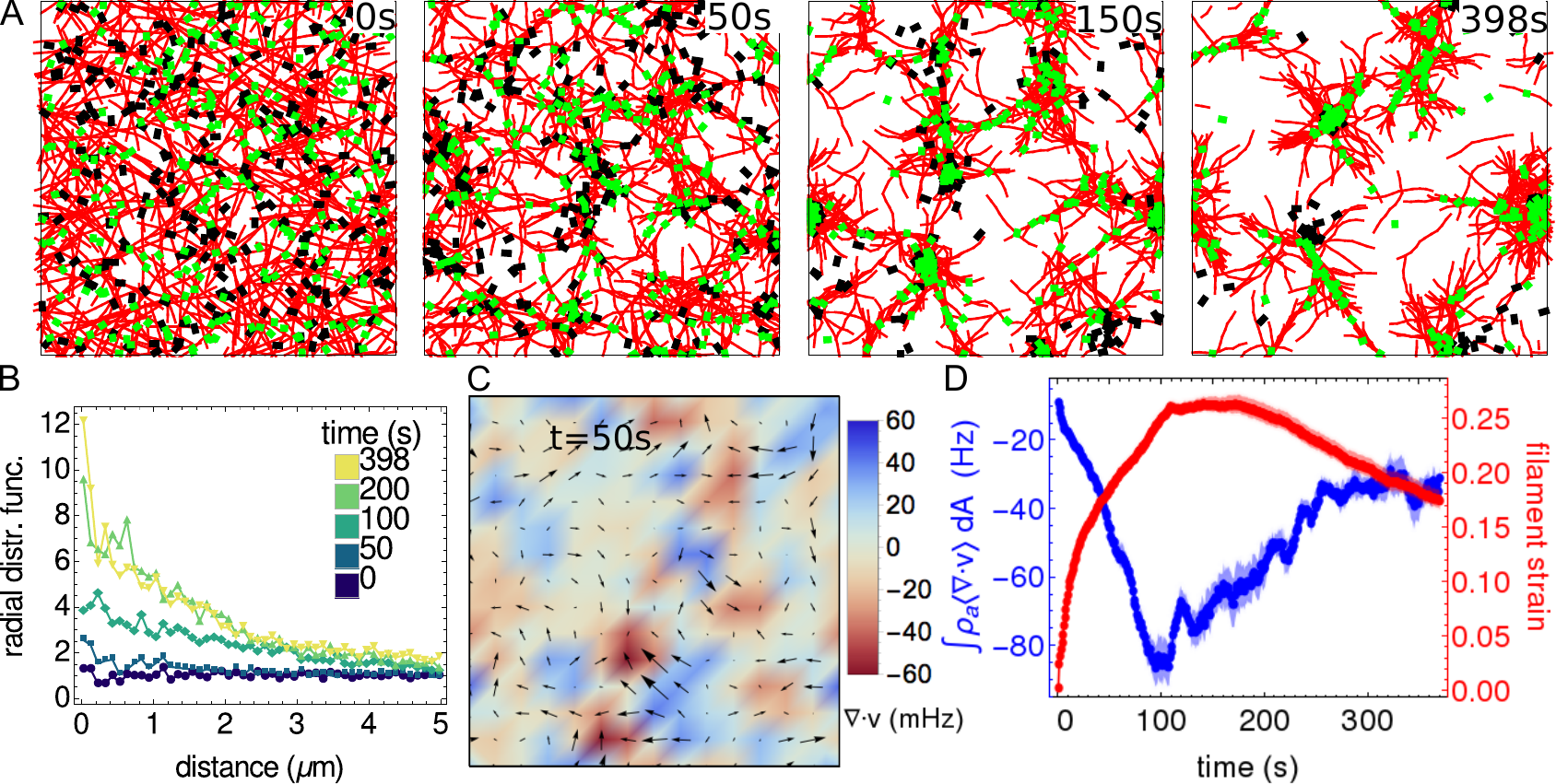}
  \caption{%
  \label{fig:network}%
  Contractility of a crosslinked filament network driven by motors. 
  Filaments are red, motors are black, and crosslinkers are green. 
  (A) Network configurations at $t=0$, $50$, $150$, and $398$ s. 
  While all filaments are shown, only $10\%$ of crosslinkers and $50\%$ of motors are shown for clarity. 
  (B) Radial distribution function at  frames corresponding to (A).   
  (C) 
  Quantification of the motion at $t=50$ s. Arrows (directions and sizes) indicate the filament-bead velocity field generated by the procedure in \Cref{app:div}. 
  Colors map the corresponding divergence. 
  (D) The density weighted divergence (blue; ${dA}=(1\ \mu$m$)^2$) and average filament strain (red) of actin filaments for contractile networks.
  Dark lines for both curves shows the mean $\mu(t)$ of these results at each time $t$ over $N=20$ simulations. 
  Shaded areas show the standard error of the mean $\mu(t)\pm\sigma(t)/\sqrt{N}$ where $\sigma(t)$ is the standard deviation.
 }
\end{figure}

When motors, crosslinkers, and filaments are combined into a single assembly, simulated networks contract.
The structure and dynamics of these networks exhibits a rich dependence on motor and crosslinker densities, binding/unbinding kinetics, and stiffness parameters.
Here, we show one illustrative example to demonstrate that our model reproduces actomyosin contractility for a reasonable choice of parameters (\Cref{fig:network}A). 
The network is initialized by randomly orienting $500$ filaments, each $10\ \mu$m long, within a $50\ \mu\textrm{m} \times 50\ \mu\textrm{m}$ simulation cell.
We distribute $0.15 \ \mu$m long crosslinkers throughout the simulation cell at a density of $1\ \mu\textrm{m}^{-2}$, and $0.5\ \mu$m long motor oligomers at a density of $0.2\ \mu\textrm{m}^{-2}$. 
As the simulation evolves, the actin density becomes more heterogenous as motors condense actin filaments into dense disordered aggregates. 
This density heterogeneity can be quantified by the radial distribution function of actin filaments, $g(r) = P(r)/(2\pi r\delta r \rho_f)$, where $P(r)$ is the probability that two filaments are separated by a distance $r$, $\delta r=0.1\ \mu$m is the spatial bin size, and $\rho_f$ is the filament density. 
As shown by \Cref{fig:network}B, $g(r)\approx1$ at $t=0$ for all $r$ as the actin filaments are homogeneously distributed. 
However, over time it becomes more peaked at lower separation distances between filaments, indicating filament aggregation.

To measure the contractile activity of the network, we evaluate the divergence of its velocity field. 
This is done by calculating the velocity of each of the actin beads, followed by a grid-based interpolation of a velocity vector field from those values (black arrows in \Cref{fig:network}C; interpolation scheme described in \Cref{app:div}). 
One can then evaluate the divergence $\nabla \cdot \vec{v}$ of the interpolated field at every spatial location (color of \Cref{fig:network}C). 
Since there is no flux of actin into the simulation box, the total divergence of the flow field is zero at all times (i.e., $\int{\left(\nabla\cdot \vec{v}\right){dA}} =0$). 
Therefore, we weight the divergence of each patch of the network by its local density and measure $\int{\rho_a\langle\nabla\cdot\vec{v}\rangle{dA}}$ where $\rho_a=n_a/{dA}$ is the number density of actin beads and $\langle\nabla\cdot\vec{v}\rangle$ is the average actin divergence in the patch of size ${dA}$. 
As shown in \Cref{app:div}, this order parameter shows consistent behavior for small patches ($dA\le(10\ \mu$m$)^2$) and a range of step sizes ($h\le20$ s) for the velocity calculation (\Cref{eqn:v(t)}).  
We also measure the average filament strain $\overline{\Delta s}$, in the network, where 
\begin{equation} 
  \Delta s = \left(1 - \frac{|\vec{r}_N-\vec{r}_0|}{\sum_{i=1}^N{|\vec{r}_i-\vec{r}_{i-1}|}}\right),
  \label{eqn:fil_strain}
 \end{equation} 
$\vec{r}_i$ is the position of the $i^{th}$ bead on an $(N+1)$-bead filament, and the bar denotes an average over all filaments.  
\Cref{fig:network}D shows the results of measuring network divergence and filament strain from $20$ simulations with the same parameter choices as in \Cref{fig:network}A, but with different random number seeds. 
The divergence measurement (blue) shows that the network is contractile, since the density weighted divergence is negative, and its shape echoes the experimental results in \cite{murrell2014}, where the magnitude of contractility decreases to a minimum before plateauing.  
The filament strain measurement (red) shows that as the network is contracting, individual filaments are buckling. 
This supports the notion that the mechanism behind contractility in disordered actomyosin networks is actin filament buckling \cite{lenz2012,murrell2012}.
We note that, while the parameterization of motors that we used for the motility assays yields contractile networks (\Cref{fig:div_supp}), using a lower value of $k_m^{off}$ resulted in kinetics closer to those observed in experiment \cite{murrell2014}.
This improvement with higher motor affinity may reflect differences in the number of participating motor heads in contractility and motility assays.

\section{Conclusion}
In this paper, we have introduced an agent-based modeling framework that can accurately and efficiently simulate active networks of filaments, motors, and crosslinkers to aid in the interpretation and design of experiments on cytoskeletal materials and synthetic analogs.  
While our focus here has been on selecting parameters that are representative of the actin cytoskeleton, we expect that this framework can be adapted to treating other active polymer assemblies as well, such as microtubule-kinesin-dynein networks.  
We demonstrated that the model gives rise to both qualitative and quantitative trends for structure and dynamics observed in experiments and provides experimentally testable predictions.  
Specifically, we reproduced the experimentally observed and theoretically described fluctuation statistics of actin filaments.
We also captured strain stiffening scalings and predicted how network elasticity can potentially be tuned via crosslinker stiffness. 
We modeled sliding filament assays and determined specific system parameters that lead to the crossover from transversely diffusive to longitudinally processive motion first predicted in \cite{duke1995}.
In separate studies, we use our model to explore the phase space of various network structures and the dynamics that lead to them \cite{stam2017}.

While our model captures many experimental observations, we simplified certain features to limit both  computational cost and model complexity.  First, the structure of myosin minifilaments is significantly more complex than a two-headed
spring. As mentioned, minifilaments have dozens of heads, which allows them to attach to more than two filaments simultaneously, significantly increasing local network elasticity
\cite{linsmeier2016} and enabling more complex motor dynamics \cite{scholz2016}.
Second, filaments do not polymerize, depolymerize, or
sever in the simulations; it is clear, however, that recycling of actin monomers, actin treadmilling and, to a lesser degree, filament severing 
play important roles in contraction and shape formation \cite{wilson2010, murrell2012}. Third, our simulations are restricted to 2D, without steric or hydrodynamic interactions.  
This can play a role in motility assays, for example, where at high actin densities, actin filaments organize into polar patterns with characteristic autocorrelation times \cite{schaller2010}.
It would be valuable to make the model a progressively more faithful representation of reality in the future to better understand how each of these choices impacts the behavior of the model and in turn the implications for the associated physics.

\section{Author Contributions} 
S.L.F., S.B., G.M.H., and A.R.D. designed the research.  S.L.F. designed and implemented the software, and executed the calculations.  S.B. provided a prototype program. 
S.L.F., S.B., G.M.H., and A.R.D. wrote the paper. 

\section{Acknowledgements}  
We thank M. Gardel, J. Weare, C. Matthews, E. Thiede, F. Nedelec, F.C. Mackintosh, and M. Murrell for helpful conversations. 
We thank C. Tung, J. Harder, and S. Mallory for critical readings of the manuscript.
This research was supported in part by the University of Chicago Materials Research Science and Engineering Center (NSF Grant No. 1420709). 
S.L.F. was supported by the DoD through the NDSEG Program. 
G.M.H. was supported by an NIH Ruth L. Kirschstein NRSA award (1F32GM113415-01).
S.B. acknowledges support from the Institute for the Physics of Living Systems at the University College London.

\section{Supporting Citations}
Reference \cite{scipy} appears in the Supporting Material. 

\bibliographystyle{biophysj}
\bibliography{actosim}
\newpage
\beginsupplement
\title{A versatile framework for simulating the dynamic mechanical structure of cytoskeletal networks: Supporting Material}
\author{S. L. Freedman, S. Banerjee, G. M. Hocky, A. R. Dinner}
\maketitle
 
\section{Calculation of crosslinker head position during binding and unbinding}\label{app:detbal}
In this section, we describe how we update the binding state ($I_{1(2)}$ in \Cref{eqn:uxl}) and position ($\vec{r}_{1(2)}$) of a crosslinker head.  The binding states and positions of the two heads of a crosslinker are coupled only through the potential energy (\Cref{eqn:uxl}).

\begin{figure}[H]
  \centering
  \includegraphics[scale=0.8]{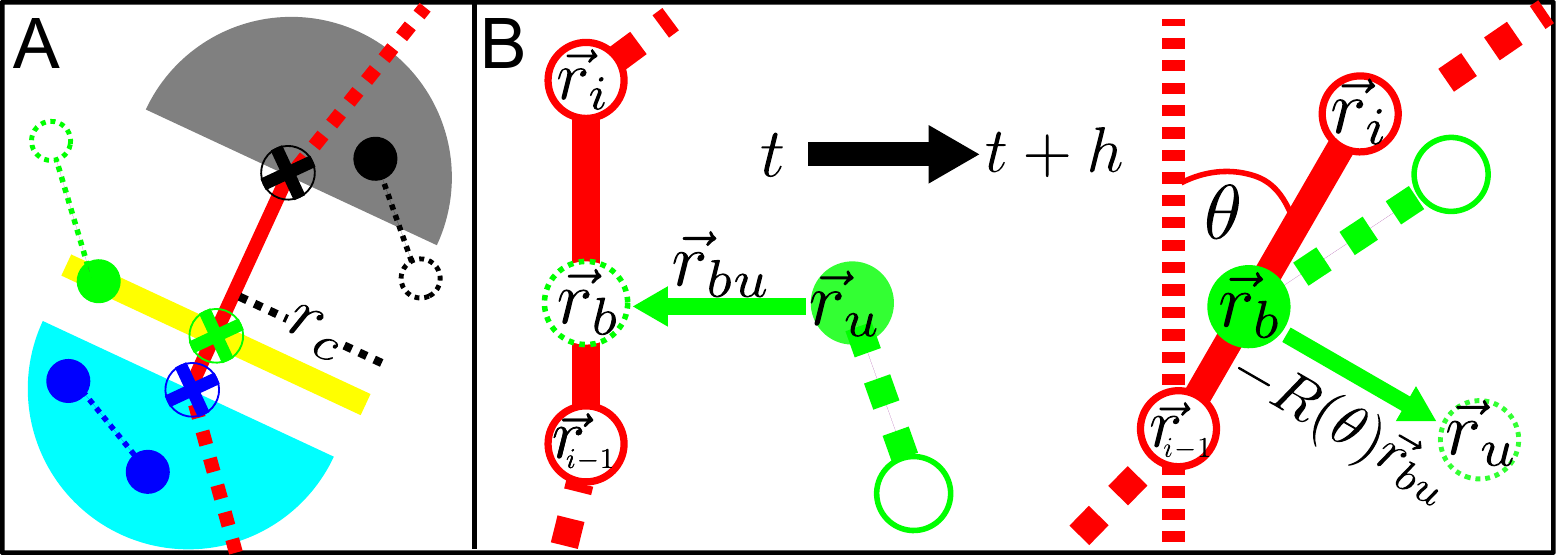}
  \caption{
  Position of crosslinker head upon binding or unbinding.
  (A) Any crosslinker head in the aqua, yellow, and gray areas (such as the filled blue, green, and black circles) can bind to the blue, green, and black binding points (circles with crosses), respectively. 
  (B) The process by which a crosslinker generates an unbinding point ($r_u$) at time $t+h$ using its original displacement at time $t$ when it snapped to the binding point $r_b$. 
}
  \label{fig:detbal}
\end{figure}

We first discuss binding.  An unbound crosslinker head with position $\vec{r}_u$ can attempt to bind to the closest point on each nearby filament link.
Let $\vec{l}_i=\vec{r}_{i}-\vec{r}_{i-1}$, where $\vec{r}_i$ is the position of the $i^{th}$ bead on the filament to which the link belongs.  Then, we propose a bound state with binding point
\begin{equation} 
  \vec{r}_b=
  \begin{cases}
    \vec{r}_{i-1} & |\vec{l}_i|=0\textrm{ or } p\le0\\
    \vec{r}_{i} & p\ge1\\
    \vec{r}_{i-1}+p\vec{l}_i&\textrm{otherwise}
  \end{cases}
    \label{eqn:rb}
\end{equation}
where $p=(\vec{r}_u-\vec{r}_i)\cdot\vec{l}_i$.
\Cref{eqn:rb} can be interpreted easily in a reference frame in which $\vec{l}_i$ is oriented vertically (\Cref{fig:detbal}A): if $\vec{r}_u$ is below the link, $\vec{r}_b=\vec{r}_{i-1}$; if it is above the filament then $\vec{r}_b=\vec{r}_{i}$; otherwise $\vec{r}_b$ is the intersection of $\vec{l}_i$ with the line perpendicular to $\vec{l}_i$ that passes through $\vec{r}_u$.
If $|\vec{r}_b-\vec{r}_u|<r_c$, the changes in binding state and position are accepted with probability $(k_{xl}^{on}\Delta t)P_{xl,i}^{off\rightarrow on}$ (see main text, \nameref{secn:crosslinkers}).

For unbinding, we do the following.  At the time of binding ($t$), we record the displacement vector, $\vec{r}_{bu}=\vec{r}_b(t)-\vec{r}_u(t)$, and the vector connecting the ends of the filament link, $\vec{l_i}(t)=\vec{r}_{i}(t)-\vec{r}_{i-1}(t)$.  At the time that we attempt unbinding ($t+h$), we determine the angle of rotation of the filament link:
\begin{equation}
\theta=\arccos{\left(\frac{\vec{l}_i(t)\cdot\vec{l}_i(t+h)}{|\vec{l}_i(t)||\vec{l}_i(t+h)|}\right)}.
\end{equation}
Then, the position to which the crosslinker head tries to jump is 
\begin{equation}
\vec{r}_u(t+h)=\vec{r}_b(t+h)-
\begin{pmatrix}
    \cos{(\theta)} & -\sin{(\theta)}\\
    \sin{(\theta)} & \hspace*{1em}\cos{(\theta)}
  \end{pmatrix}
\vec{r}_{bu}(t)
\end{equation} 
as shown in \Cref{fig:detbal}B.
This jump is accepted with probability $(k^{off}_{xl}\Delta t)P_{xl,i}^{on\rightarrow off}$.
The motivation for this scheme is that it ensures that a head that jumps onto (off) a filament link returns to its original position if it unbinds (rebinds) immediately.  Detailed balanced consistent with \Cref{eqn:uxl} can thus be satisfied through the acceptance probabilities $(k^{on}_{xl}\Delta t)P_{xl,i}^{off\rightarrow on}$ and $(k^{off}_{xl}\Delta t)P_{xl,i}^{on\rightarrow off}$.

\section{Relaxation times scales}\label{app:fil}
In this section, we present data on filament and network time scales that inform our choices of sampling frequencies.

\subsection{Decorrelation of filament angles}

\begin{figure}[H]
  \centering
  \includegraphics[scale=1.2]{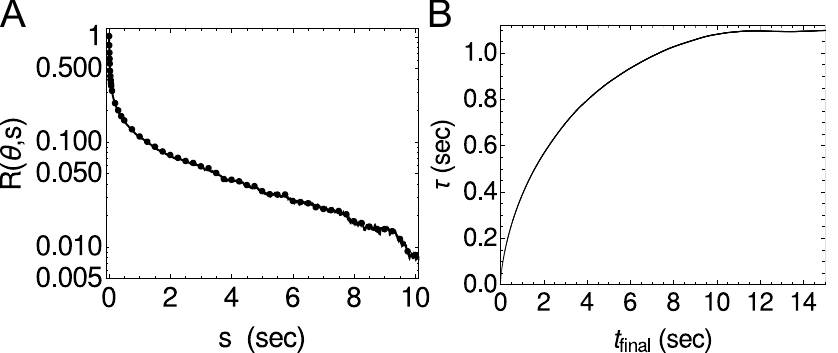}
  \caption{ \label{fig:pl_supp_decor}Estimation of the characteristic decorrelation time for persistence length measurements. 
  (A) Decorrelation of angles between filament links for a $21$ bead filament with $k_a=1$ pN$/\mu$m, $l_a=1\ \mu$m, and $\kappa_B=0.068$ pN$\mu$m$^2$. 
  (B) Measurement of the integrated autocorrelation time $\tau$ for different values of the cutoff time $t_{final}$.  
 }
\end{figure}

The evaluations of persistence length in \nameref{secn:pl} in the main text average over independent configurations of filaments. 
To determine the amount of time between independent configurations in 
a trajectory of a single filament, we evaluated the integrated autocorrelation time of the angles $\theta_i$ for $i\in [2\ldots 20]$  between links along a 
$21$ bead filament. \Cref{fig:pl_supp_decor}A shows the autocorrelation 
\begin{equation}
R(\theta,s)=\frac{\langle \theta(t)\theta(t+s)\rangle-\langle\theta(t)\rangle^2}{\langle \theta(t)^2\rangle-\langle\theta(t)\rangle^2}
\label{eqn:auto}
\end{equation}
where $s$ is the time between realizations and the angle brackets represent an average over all $19$ angles and all $1900$ saved configurations.  
\Cref{fig:pl_supp_decor}B shows the integrated autocorrelation time $\tau$ as a function of the simulation cutoff time $t_{final}$, where
\begin{equation}
\tau(\theta) =\int_{0}^{t_{final}}{R(\theta,s)ds}.
\label{eqn:iat}
\end{equation}
For all choices of $t_{final}$, $\tau<2$ s and therefore configurations that are separated by at least $2$ s should be  independent realizations with respect to angles between subsequent filament links. 

\subsection{Shear relaxation times}\label{app:shear}
\begin{figure}[H]
  \centering
  \includegraphics[scale=1.2]{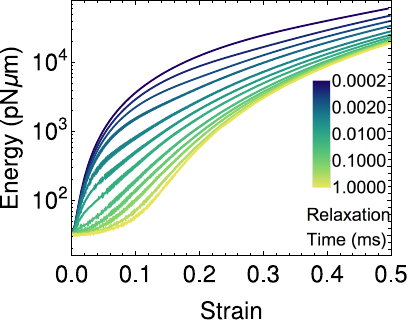}
  \caption{\label{fig:elas_supp}
    Total potential energy as a function of strain for various relaxation
    times. Simulation parameters, are otherwise identical to the shear simulations
  in the main text.} 
 \end{figure}

One extra parameter that must be set for shear simulations is the relaxation time ($t_{relax}$)---i.e., the minimum time between strain steps for responses to be history independent. 
We probed this question computationally by determining if the parameter of interest (total potential energy of filaments and crosslinkers) varied significantly for different periods of relaxation between steps of $\Delta \gamma=0.001$. 
\Cref{fig:elas_supp} shows that while very small $t_{relax}$ values do yield higher energies at equivalent strains, as $t_{relax}$ is increased, the curves collapse for identical strains. 
In the shear simulations in the main text (\nameref{secn:shear}), $t_{relax}=1$ ms (yellow curve). 

\section{Comparison with CytoSim}\label{app:cytosim}
Cyotsim is a freely available C++ software package developed to simulate active polymer networks and described in \citeSupp{nedelec2007S}. 
While AFiNeS shares many of the same features, for clarity we enumerate the technical differences. 
\begin{itemize}
\item{\textbf{The filament model.} 
AFiNeS uses a bead spring chain and CytoSim uses a chain constrained via Lagrange multipliers. }
\item{\textbf{Attachment of motors and crosslinkers.} 
CytoSim uses a continuous-time Monte Carlo procedure (the Gillespie algorithm \citeSupp{gillespie1977S}) to calculate when a motor should attempt attachment to a filament, while AFiNeS attempts with the probability computed for each discrete timestep of fixed duration. 
In Cytosim, the attachment of a motor to a filament is not dependent on the distance from the filament, other than that it must be below a threshold, whereas in AFiNeS, a closer motor has a higher probability of attachment, due to detailed balance considerations.}
\item{\textbf{Detachment of motors and crosslinkers.} 
CytoSim has a force dependent detachment of crosslinkers.
This was not a necessary detail to reproduce the benchmarks shown in the results section, and detailed balance would require altering the motor and crosslinker dynamics, so we have not included it in the present version.
We plan in the future to understand how this detail effects cytoskeletal networks in general and add it as an option to AFiNeS.}
\item{\textbf{Capabilities present in one and not the other.} 
AFiNeS implements network shearing. 
CytoSim implements filament polymerization and depolymerization, microtubule asters, and spherical geometries. 
}
\end{itemize}

\begin{figure}[H]
  \centering
  \includegraphics[scale=1.2]{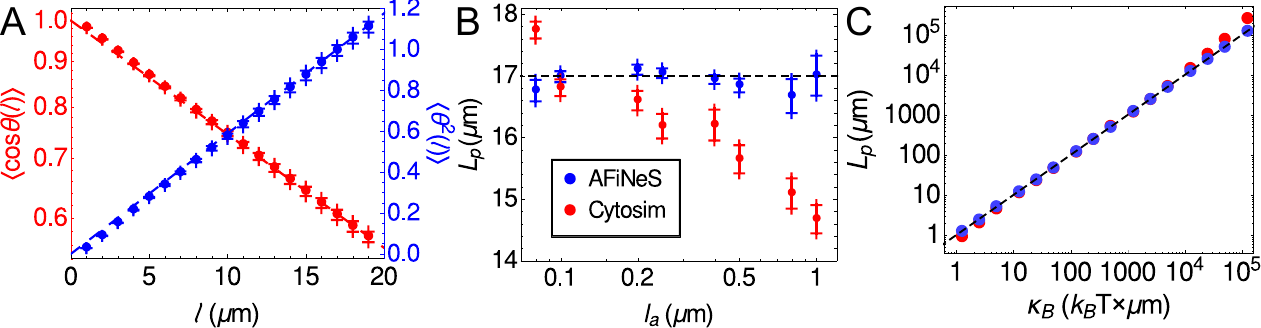}
  \caption{Measurements of persistence length for CytoSim filaments (red) compared with the same measurements for AFiNeS (blue).
        (A) Cosine correlation function and $\Delta \theta^2$ correlation function for $20$ CytoSim fibers with $L_p=17\ \mu$m fluctuating for $2000$ seconds. 
        See Section 4.1 of the main text for details. 
      (B) Measurement of $L_p$ as function of segment length, $l_a$, using the fit to the first 5 data points of $\langle \Delta\theta^2\rangle$ in (A). 
        (C) Measurement of $L_p$ as a function of input bending modulus for CytoSim and AFiNes.  Colors are the same as panel B.
      }
  \label{fig:cs_lp}
\end{figure}

To compare the two packages, we have used CytoSim to run the benchmarks associated with filament fluctuations (\Cref{fig:cs_lp}) and motility assays (\Cref{fig:cs_motility}, below). 
For the filament fluctuation benchmarks, shown in \Cref{fig:cs_lp}, we find that while CytoSim is able to yield nearly the correct persistence length of filaments, at long segment lengths it performs worse than AFiNeS, perhaps because it uses linearized versions of the angle forces \cite{nedelec2007S}.
\section{Parsing the energy in sheared networks}\label{app:sheareng}
\begin{figure}[H]
  \centering
  \includegraphics[scale=1.2]{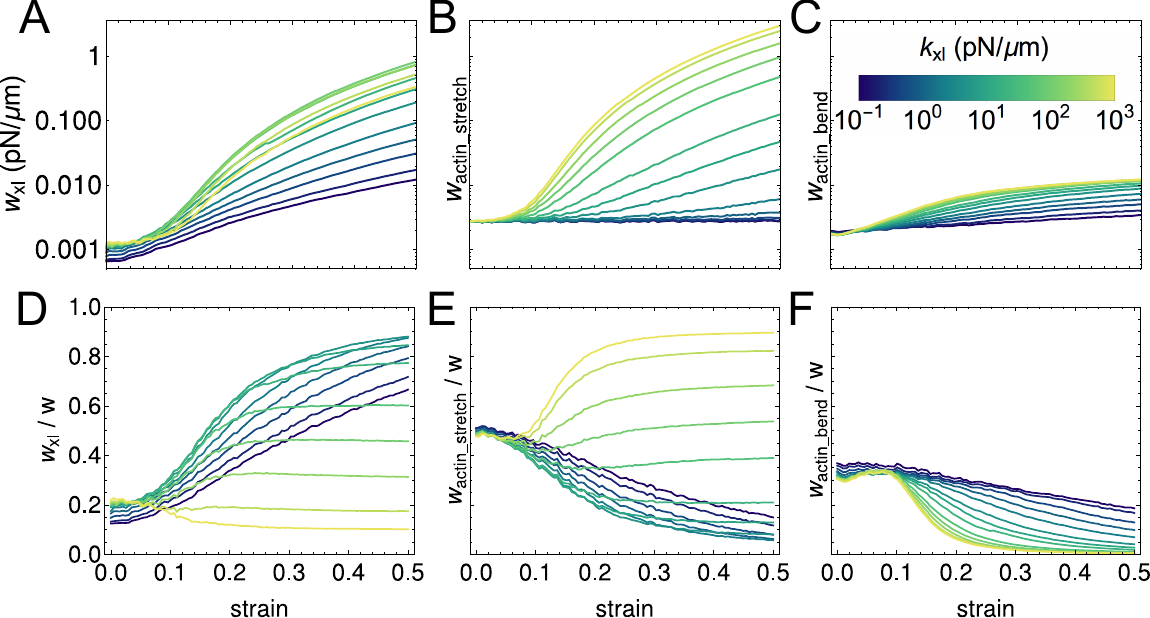}
  \caption{Absolute (A-C) and relative (D-F) energy contributions from crosslinkers stretching (A, D), filaments stretching (B, E), and filaments bending (C, F) for the sheared network discussed in \nameref{secn:shear}. }
  \label{fig:engs}
\end{figure}

To further examine the source of the energy scalings shown in \Cref{fig:stress}D, we measure the fraction of the total energy density $w$ from each of its sources in the network, the stretching energy of filaments, the stretching energy of crosslinkers, and the bending energy of filaments, as shown in \Cref{fig:engs}.
In general, we find that shearing the network stretches and bends actin filaments, and also stretches crosslinkers, as in \Cref{fig:engs}A-C. 
\Cref{fig:engs}B-C show that, as crosslinkers become more stiff, more of the energy from the strain is concentrated on the filaments. 
\Cref{fig:engs}A shows that for crosslinkers, the trend is not monotonic. 
When $k_{xl}<100$ pN$/\mu$m, increasing crosslinker stiffness results in more energy in the crosslinkers, and in this regime, the scaling of $w(\gamma)$ increases monotonically. 
However, for $k_{xl}\ge100$ pN$/\mu$m, the trend reverses, and the strain energy density concentrates on the filaments more than the crosslinkers, as seen in \Cref{fig:engs}D-E.
In this regime, the scaling of $w(\gamma)$ plateaus near the value 3.5, reflecting the prediction for the differential shear modulus in a strain controlled rheology experiment, $G={d^2w}/{d\gamma^2}\propto\gamma^{3/2}$ \citeSupp{gardel2004S}.

\section{Comparison with Cytosim for motility assays}\label{app:cytosim_mot}
\begin{figure}[H]
  \centering
  \includegraphics{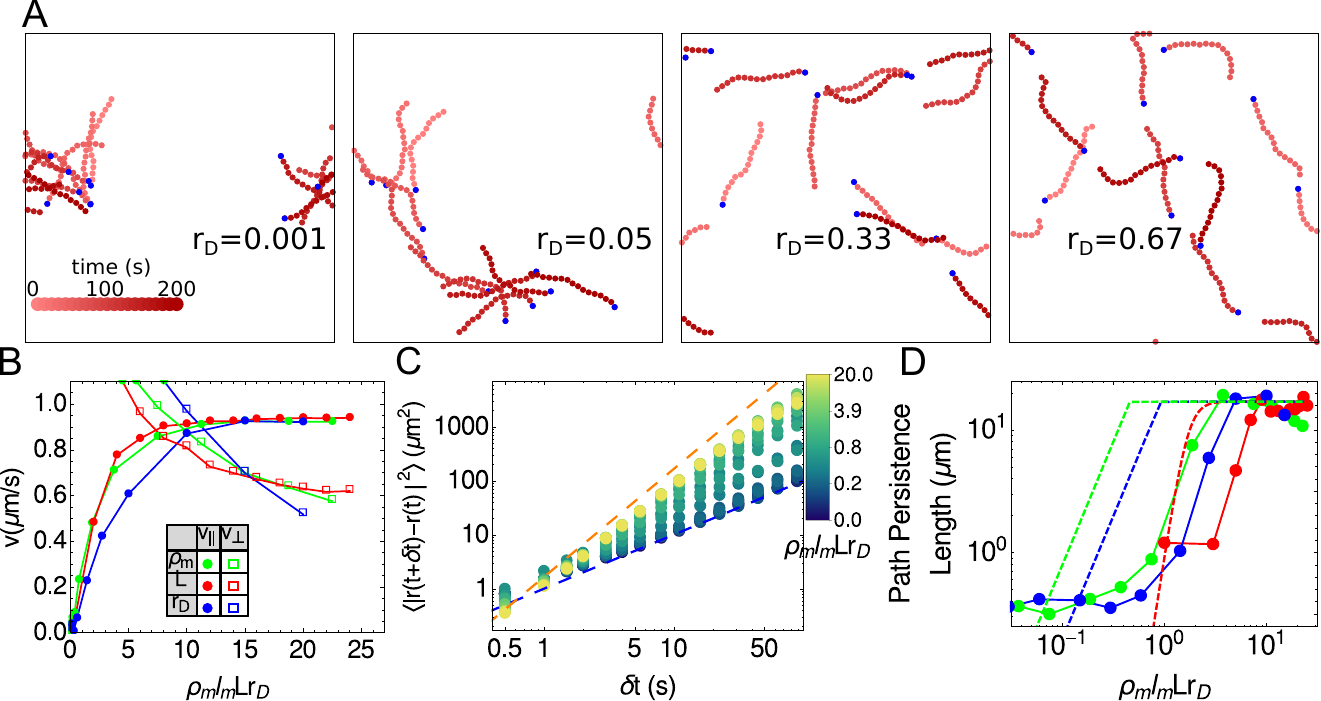}
  \caption{Motility measurements at varying motor density, filament length, and duty ratio generated using CytoSim. 
    For a detailed description of this calculation see main text, \nameref{secn:motility}.} 
  \label{fig:cs_motility}
\end{figure}

We also used Cytosim to simulate the motility assays described in the main text (\nameref{secn:motility}).
The results, shown in \Cref{fig:cs_motility}, are generally congruent with the results from AFiNeS in \Cref{fig:motility}.
We find that increasing motor density, filament length, and duty ratio increase longitudinal motion and decrease transverse motion of the filament (\Cref{fig:cs_motility}B), and makes the filament move more ballistically (\Cref{fig:cs_motility}C).
Furthermore, the path persistence length plots (\Cref{fig:cs_motility}D) are nearly identical to the measurements obtained using AFiNES.
Thus, it is reassuring that the two models agree to this extent despite the differences in filament and binding implementations. 

\section{Procedure for quantifying contractility}\label{app:div}
An actin assay can be considered contractile if it has regions to which most of the actin aggregates.  
In an experiment with a limited field of view, the net flux of actin into the field of view is positive when the system is contractile.  
This flux corresponds mathematically to a negative value for the integral of the divergence of the velocity field over the area \citeSupp{murrell2012S, murrell2014S}. 
However, in our simulations, all particles' positions are known and there is no flux of material into or out of the simulation region owing to the periodic boundary condition. 
Thus the total divergence obtained by integrating over the simulation box must be zero. 
Nevertheless, we can still compute the density-weighted divergence to quantify contractility, as we now describe. 

\begin{figure}[H] 
    \centering
    \includegraphics[width=6.75in]{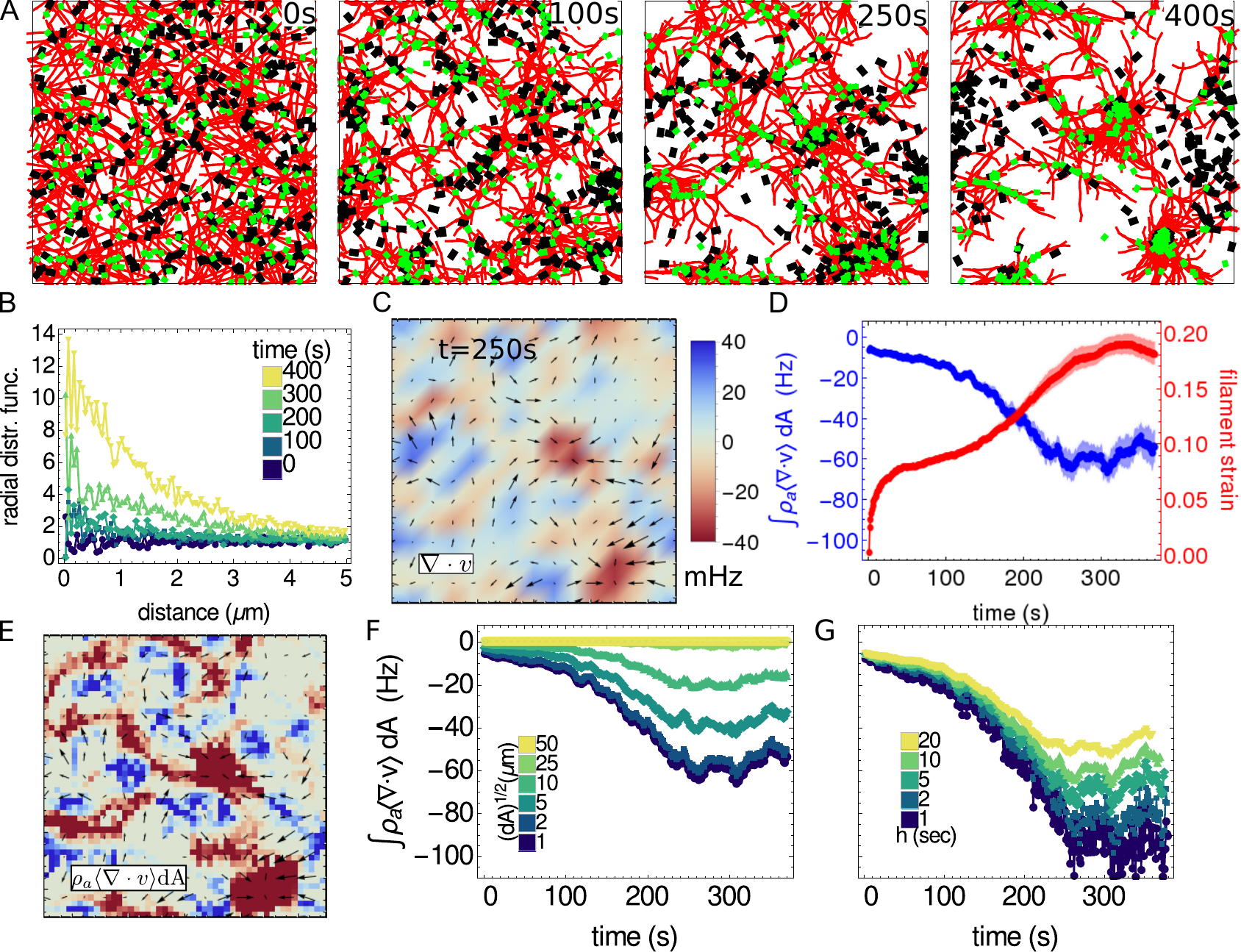}
  \caption{%
  \label{fig:div_supp}%
  Calculation of density weighted divergence for a simulated contractile actomyosin network.
  (A-D) Identical to \Cref{fig:network}, but with $k_m^{off}=10$ s$^{-1}$, $k_m^{end}=1$ s$^{-1}$, and $\rho_m=1\ \mu$m$^{-2}$. 
  In (A), all filaments and $10\%$ of motors and crosslinkers are shown. 
  (E) Same as (C), but the color is weighted by the actin density $\rho_a$.
  (F) Dependence of the density weighted divergence on the patch size used for integration, ${dA}={dxdy}$, with $h=10$ s.  
  (G) Dependence of the density weighted divergence on the time scale $h$ used in calculating the velocity of actin $v$ in \Cref{eqn:v(t)} with ${dx}={dy}=1\ \mu$m. 
    }
 \end{figure}

To ensure that the divergence is well-defined at all points, we first interpolate a continuous velocity field.
When the data are experimental images, the velocity field is determined using Particle Image Velocimetry (PIV).
Here, we take a similar approach, with the advantage that positions of actin beads are a direct output of the simulation, analogous to tracer particles in experiments. 
To this end, for each filament bead $i$ with position $\vec{r}_i(t)$ at time $t$, we calculate the velocity by forward finite difference: 
\begin{equation}
  \vec{v}_i(\vec{r}_i,t)=\frac{\vec{r}_i(t+h)-\vec{r}_i(t)}{h},
  \label{eqn:v(t)}
\end{equation} 
where $h$ is a suitable amount of time to characterize motion.
We calculate the average velocity of each $(5\ \mu\textrm{m})^2$ bin.
Similarly to PIV, we lower the noise further by setting a threshold, and only consider bins with at least $n$ actin beads.
We then interpolate the bin values with Gaussian radial basis functions (RBFs):
\begin{equation}
\vec{v}(\vec{r})=\sum_{k=1}^M{\vec{w}_{k}e^{-\left(|\vec{r}-\vec{r}_k|/\epsilon\right)^2}}
\label{eqn:interp}
\end{equation}
where $M$ is the number of bins with at least $n$ actin beads, $\epsilon$ is a constant related to the width of the Gaussian RBFs, and $\vec{w}_{k}$ are their weights. 
The optimal value for $\epsilon$ is generally close to the value of the average distance between RBFs \citeSupp{scipyS}; we found $\epsilon=5\ \mu$m and a threshold of $n=10$ yielded a robust interpolation across many different actin structures. 
We use the scipy.interpolate.Rbf Python package to determine the weights \citeSupp{scipyS}. 
We calculate the divergence of the resulting field $dv_x(\vec{r})/dx+dv_y(\vec{r})/dy$ by using finite difference approximations for the derivatives of \Cref{eqn:interp}.
Examples of this velocity field and the local divergence are shown in \Cref{fig:network}C and \Cref{fig:div_supp}C.

As noted above, given $\nabla\cdot\vec{v}$, we quantify the contractility by the density weighted divergence, $\int{\rho_a\langle\nabla\cdot \vec{v}\rangle dA}$.  
In \Cref{fig:div_supp}E we show an example where the density weighting has the effect of significantly increasing the magnitude of the areas with negative divergence. 
To understand how the contractility varies with length scale, we replace the integral with the sum over square regions
\begin{equation}
\sum_k{\rho_a(\vec{r}_k)\langle\nabla\cdot \vec{v}\rangle_k dA}
\end{equation}  
and vary the size of the regions, $dA=dxdy$ (\Cref{fig:div_supp}F).
For the maximum size $dA=(50\ \mu\textrm{m})^2$ (yellow curve), the density weighted divergence fluctuates around $0$ as expected from the zero actin flux.
However for region sizes $dA\le(10\ \mu$m$)^2$, the values are consistently negative, indicating contractility; the curves decrease to a minimum before plateauing closer to $0$, as seen in experiment \citeSupp{murrell2014S}. 
We also show, in \Cref{fig:div_supp}G, that the trend of this order parameter is independent of the time scale $h$ used to calculate the velocity in \Cref{eqn:v(t)}.

\bibliographystyleSupp{biophysj}
\bibliographySupp{actosimSupp}

\end{document}